\documentclass[aps,onecolumn,superscriptaddress,groupedaddress,notitlepage]{revtex4-1}  
  \usepackage[bookmarks=false,colorlinks]{hyperref}

\makeatletter
\renewcommand{\p@subsection}{}
\renewcommand{\p@subsubsection}{}
\makeatother

\usepackage[T1]{fontenc}

\usepackage{amssymb,amsmath,amsfonts}
\usepackage{epsfig}
\usepackage{graphicx}
\usepackage{epstopdf}
\usepackage{caption}
\usepackage{subcaption}
\usepackage{braket}
\usepackage{amscd}
\usepackage{stmaryrd}
\usepackage{amsthm}
\usepackage{latexsym}
\usepackage{amsbsy}
\usepackage[english]{babel}
\usepackage{psfrag}
\usepackage{tabularx}
\usepackage{bbm}
\usepackage{overpic}
\usepackage{array}
\usepackage{rotating}
\usepackage{color}
\usepackage{comment}

\def\be{\begin{equation}}
\def\ee{\end{equation}}
\def\bea{\begin{eqnarray}}
\def\eea{\end{eqnarray}}

\def\tr{\mbox{tr}}

\newcommand{\eq}[1]{(\ref{#1})}

\newcommand{\la}{\label}
\newcommand{\ba}{\begin{align}}

\def\12{\frac{1}{2}}

\newcommand{\en}{\end{align}}

\usepackage{color}

\theoremstyle{plain}
\newtheorem{thm}{Theorem}

\newtheorem{prop}{Proposition}
\newtheorem{lemma}{Lemma}

\theoremstyle{definition}
\newtheorem{defn}[thm]{Definition} 

\begin{document}


\title{Random Lindblad Dynamics}
\author{Tankut Can}
\affiliation{Initiative for the Theoretical Sciences, The Graduate Center, CUNY, New York, NY 10016, USA}

\begin{abstract}

We study the mixing behavior of random Lindblad generators with no symmetries, using the dynamical map or propagator of the dissipative evolution. In particular, we determine the long-time behavior of a dissipative form factor, which is the trace of the propagator, and use this as a diagnostic for the existence or absence of a spectral gap in the distribution of eigenvalues of the Lindblad generator. We find that simple generators with a single jump operator are slowly mixing, and relax algebraically in time, due to the closing of the spectral gap in the thermodynamic limit. Introducing additional jump operators or a Hamiltonian opens up a spectral gap which remains finite in the thermodynamic limit, leading to exponential relaxation and thus rapid mixing. We use the method of moments and introduce a novel diagrammatic expansion to determine exactly the form factor to leading order in Hilbert space dimension $N$. We also present numerical support for our main results. 

\end{abstract}


\maketitle 
\tableofcontents

\section{Introduction}


 


A foundational problem in statistical mechanics concerns the existence, or lack thereof, of thermalization. Indeed, Boltzmann was among the first to address the question of whether many-body systems reach equilibrium \cite{Boltzmann1872}. Closely related to this issue is the question of {\it how} non-equilibrium systems approach equilibrium, in particular the dynamics of relaxation. 
 

Since Boltzmann, the conditions under which thermalization is achieved have been clarified considerably, and there now seems to be a clear picture of the roles of chaos, ergodicity, and integrability in determining whether thermalization occurs in classical and quantum systems. However, there seems to be much left to say about the story of non-equilibrium relaxation, and characterizing for instance the typical time scales involved in the approach to equilibrium \cite{Goldstein2013}, and the effects of metastability \cite{Lesanovsky2013,Macieszczak2016}. 

Thermalization in classical systems is intimately tied to non-integrability and chaoticity of the microscopic dynamics. The quantum manifestation of chaos is the appearance of a universal random matrix theory (RMT) description of late-time dynamics and spectral statistics  \cite{Berry1985,Bohigas1984}. The eigenstate thermalization hypothesis (ETH) provides the bridge between the universal RMT behavior and thermalization in quantum statistical mechanics \cite{Srednicki1999,Rigol2008,Dymarsky2018a,DAlessio2016,Schiulaz2018}. 

While it is valuable to clarify this story for isolated systems, they are still rather the exception in nature. It is natural to wonder if driven and dissipative systems also exhibit universal dynamical or steady state properties. Scattering theory has proven to be a fruitful approach to addressing the question of universality in open quantum systems. A conjecture for open chaotic scattering involving the distribution of phase shifts was proposed in \cite{Blumel1990}, as the open analog of the Bohigas-Giannoni-Schmit conjecture for level statistics in chaotic billiards \cite{Bohigas1984}. In the scattering problem, the object of interest is the scattering $S$ matrix whose poles are the scattering resonances. For a chaotic scattering region, the $S$ matrix is described by Dyson's circular ensembles of random unitary matrices \cite{Dyson1962a, Schomerus2016}. Furthermore, the resonance width distribution in the limit of weak coupling to the continuum is described by the Porter-Thomas distribution, which is also reproduced by random matrix theory \cite{Fyodorov1997a} (for reviews, see e.g. \cite{Schomerus2016,Fyodorov2010}). Many of these results have also received experimental support \cite{Alt1995,Kuhl2005,Kuhl2008,Kuhl2005a}. 

The dynamical signatures of quantum chaos in the presence of dissipation and decoherence have been taken up in, e.g. \cite{Zurek1994,Miller1998a,Casati1997,Waltner2018}. In the context of open chaotic scattering, the time evolution of decay functions or survival probabilities have been studied using random matrix theory \cite{Dittes1992,Harney1992,Dittes2000}.

Motivated by the success of RMT in describing the dynamical signatures of thermalization in closed quantum systems, we seek to expand this to open quantum systems by studying a random matrix theory of quantum master equations. In particular, we are interested in the approach to steady state, and the typical relaxing behavior of open quantum systems. Our results appear to be consonant with the literature on classical open chaotic scattering, suggesting that these domains can be bridged, and that a RMT description of open quantum master equations can capture universal relaxation dynamics. This discussion mainly serves to provide a larger context for our work, and we leave the precise connection to quantum chaos as a direction for future research.

The spirit of the present paper is more in line with the original approach in nuclear physics to develop a statistical description of complex quantum phenomena. A statistical description of deterministic processes can only ever have a chance of being meaningful when there are many degrees of freedom involved, allowing us to retain only the essential structure (e.g. hermitian vs non-hermitian) and symmetries (e.g. time-reversal, parity, etc. \cite{Dyson1962,Altland1997}), while eschewing complex microscopic details \cite{Dyson1962a,Wigner1955,Dyson1962}. These are then maximum entropy ensembles subject to certain constraints imposed by the physical problem of interest \cite{Balian1968}. This rule of thumb is borne out by the success of RMT in describing chaotic systems, whose motion in phase space is ergodic and highly mixing, as well as heavy nuclei which are highly complex interacting systems with many degrees of freedom. The universality of the statistical description is observed in properties which depend only on the general structure and symmetries. In developing a statistical description of non-equilibrium dissipative quantum dynamics, the ensembles we consider in this paper  are the maximum entropy ensembles of generators of the quantum dynamical semigroup, which govern the quantum Markov master equation, more familiarly known as the Lindblad equation. The general representation of Lindblad generators splits the evolution into a unitary part, governed by a Hamiltonian, and a dissipative part involving jump operators. Selecting the Hamiltonian and jump operators from a random matrix ensemble then naturally leads to a random Lindblad generator. Furthermore, one may consider discrete symmetries as in \cite{Buca2012,albert2014symmetries}, and their effect on the statistical properties of the master equation; we save this for future work, and consider here only ensembles without any additional symmetries.

We now proceed to review the relevant results from the theory of open quantum systems which we will use to define our random matrix ensembles and introduce our main object of study. We refer to \cite{Petruccione} for more details on this standard textbook material.
\smallskip

\paragraph*{Open quantum systems} Quantum master equations provide a framework to study  driven and dissipative quantum systems through the continuous time evolution of the density matrix
\begin{align}
\dot{\rho} = \mathcal{L}[\rho], \label{1}
\end{align}
where the Liouvillian $\mathcal{L}$ is in general an integral operator, non-local in time, and possibly non-linear in $\rho$. Formally integrating the master equation one can define the dynamical map $K_{t}: \mathcal{S}(\mathcal{H}) \to \mathcal{S}(\mathcal{H})$ as a map acting on the space of density matrices $\mathcal{S}(\mathcal{H})$ over the Hilbert space $\mathcal{H}$. We consider finite but large $N$-dimensional Hilbert spaces. The Liouvillian is a superoperator acting as a linear map on the  space of bounded operators $\mathcal{B}(\mathcal{H}) = \mathcal{H} \otimes \mathcal{H}^{*}$. It also has a non-unique $N^{2}\times N^{2}$ matrix representation $\mathcal{L}_{ab} = {\rm tr} (e_{a}^{\dagger} \mathcal{L}[e_{b}])$ whose specific form depends on the basis vectors $e_{a} \in \mathcal{B}(\mathcal{H})$. We use the standard basis below. 

There exists a useful operator sum representation of linear maps on the space of density matrices  due to Kraus \cite{Kraus1971} (and independently Choi \cite{Choi1975}). Such maps have come to be known as quantum operations or channels. The space of density matrices consists of complex-valued positive-definite hermitian matrices with unit trace. Any physically sensible dynamical evolution of the density matrix must preserve these properties. Kraus' theorem states that any completely positive and trace preserving (CPTP) quantum channel which takes $\rho \to \rho'$ can be represented by
\begin{align}
\rho' = \sum_{k = 1}^{N^{2}} A_{k} \rho A_{k}^{\dagger}, \quad \sum_{k} A_{k}^{\dagger}A_{k} = \mathbbm{1},\label{krauss}	
\end{align}
for some Kraus operators $A_{k}$ subject to the stated constraint. The crucial assumption which allows the explicit representation of the Kraus map is complete positivity, which ensures that every trivial extension of the map to an enlarged Hilbert space, which acts as the identity on this additional space, will preserve the positivity of the density matrix.

A subset of quantum channels are those which are parametrized by a continuous time variable and form a semigroup. A dynamical map forms a {\it quantum dynamical semigroup} (QDS) if it satisfies the following axioms
\begin{align}
(1) \,\,K_{t+s} = K_{t} K_{s}	, \quad (2)\,\, K_{0} = \mathbbm{1}, \quad  \quad t,s\ge 0.
\end{align}
Axiom (1) is equivalent to the Markov property, and implies that a QDS necessarily describes Markovian dynamics. The Markovian assumption allows us to utilize a fundamental theorem  independently discovered by Lindblad \cite{Lindblad1976} and Gorini, Kossakowski, \& Sudarshan (GKS) \cite{Gorini1976}, concerning the classification of the generators of a QDS. The generator $\mathcal{L}$ of the QDS can be written (in the ``diagonal" Lindblad basis)
\begin{align}
\mathcal{L} &= \mathcal{C}_{H} + \sum_{a = 1}^{m} \gamma_{a} \mathcal{D}_{L_{a}}.\label{lindblad}
\end{align}
We refer to the Liouvillian in this form as the Lindblad generator or superoperator, or simply the Lindbladian \footnote{Also known as the GKS-Lindblad or GKLS generator. We use generator and superoperator interchangeably when referring to the Lindbladian.}. We have split it up into two pieces: the first describes unitary Liouville-von Neumann evolution under a (Hermitian) Hamiltonian $H$ 
\begin{align}
\mathcal{C}_{H}[\rho] &= - i \left( H \rho - \rho  H \right),
\label{von-Neumann}
\end{align}
which we refer to as the Liouville-von Neumann (LvN) generator for $H$. We work mainly with the following matrix representation
\begin{align}
\mathcal{C}_{H} & = - i \left( H \otimes_{t} \mathbbm{1} - \mathbbm{1} \otimes_{t} H\right), \label{matrix-von-Neumann}
\end{align}
where $\mathbbm{1}$ is the identity matrix of dimension $N$. We have introduced the notation $\otimes_{t}$ for clarity of presentation, which is defined $A \otimes_{t} B \equiv  A \otimes B^{T}$, with $\otimes$ the conventional tensor (Kronecker) product of matrices, and $B^{T}$ denoting the matrix transpose of $B$. In this paper, we work almost exclusively with the matrix representation (\ref{matrix-von-Neumann}), and distinguish the superoperator as a map by explicitly including an argument, as in (\ref{von-Neumann}). The  dissipative part of the Lindbladian is described by a sum of {\it dissipators} $\mathcal{D}_{L}$ for the operator $L$ defined by the superoperator

\begin{align}
\mathcal{D}_{L}[\rho]  &= 2L \rho L^{\dagger} - L^{\dagger} L \rho - \rho L^{\dagger}L,
\end{align} 
which has the matrix representation
\begin{align}
\mathcal{D}_{L}  &= 2L \otimes_{t} L^{\dagger} - L^{\dagger} L \otimes_{t} \mathbbm{1} - \mathbbm{1}\otimes_{t} L^{\dagger}L.\label{dissipator}
\end{align}

The theorem imposes no constraints on the ``jump operators" $L_{a}$. The dissipative couplings $\gamma_{a}$ (also called dephasing rates), are required to be positive to ensure the convergence of the flow to a stationary state. The most general CPTP map of the Kraus form has $N^{4} - N^{2}$ independent parameters, which translates to a general dissipator consisting of $m = N^{2}- 1$ jump operators, or dissipation channels. However, the number of the jump operators, and the shape they take, is often informed by the physical problem at hand. We only consider Lindbladians with a number of dissipation channels that remains finite in the thermodynamic limit.

Sometimes, the entire dissipative contribution to the Lindblad generator is collectively referred to as the dissipator. To avoid confusion, we refer to \eq{dissipator} as the {\it simple dissipator} (also called the simple generator  \cite{Baumgartner2008}). Following the literature, we refer to the first term in \eq{dissipator} as the ``recycling term".

Given the GKS-Lindblad representation of the generator for the quantum Markov master equation, we can define an ensemble of generators by taking the Hamiltonian and jump operators to be random matrices. In this paper, we consider matrices without any additional discrete symmetries, and for simplicity assume Gaussian distributed matrix elements. Thus, we take the Hamiltonian from the Gaussian unitary ensemble (GUE) of complex Hermitian matrices. We take the jump operators from the complex Ginibre ensemble of complex matrices with no symmetries. We also present results for the simple dissipator with jump operators drawn from the GUE and normal random matrix ensembles. 

In this work, we study relaxation times under random Lindblad evolution by focusing on the dynamical map or propagator $K_{t}$. Our approach is sensitive to the important question of whether relaxation occurs exponentially fast, and is thus rapidly mixing, or algebraically in time. From this relaxation behavior we can infer the mean spectral gap in the large $N$ limit. We develop a diagrammatic approach to studying random Lindblad generators using the method of moments. Our main results are summarized in the next section.

\subsection{Basic Definitions and Main Results}

The primary object of study in this paper is the dynamical map or propagator
\begin{align}
K_{t} =  e^{ t\mathcal{L}}, \quad t\ge0,\label{propagator}
\end{align}
which forms a single parameter representation of a quantum dynamical semigroup. We take the Liouvillian to be of Lindblad form. Since we consider a finite $N$-dimensional Hilbert space, we work with the $N^{2}\times N^{2}$ matrix representation of the Lindblad superoperator, in which the propagator is just the exponential of the Lindbladian. We access asymptotic relaxation by studying the {\it dissipative form factor} (DFF) for the Lindbladian:

\begin{defn}[Dissipative Form Factor]
The dissipative form factor is defined as the ensemble averaged trace of the propagator (\ref{propagator})
	\begin{align}
F(t) = \frac{1}{N^{2}} \left\langle {\rm tr}\, K_{t} \right\rangle, \label{dff-def}
\end{align}
normalized such that $F(0) = 1$. Angular brackets indicate averaging over the relevant ensemble of Hamiltonian and/or jump operator(s).
\end{defn}

The DFF reduces to the spectral form factor (normalized by $N^{2}$) for unitary Liouville-von Neumann evolution without a dissipator. The form factor is nothing but the moment generating function for the Lindblad generator, where the {\it trace moments} are defined 
\begin{align}
M_{n} = \frac{1}{N^{2}} \left\langle {\rm tr} \, \mathcal{L}^{n} \right\rangle.
\end{align}
Here, $\mathcal{L}^{n}$ is the $n^{th}$ matrix power of the Lindblad generator, which is equivalent to the matrix representation of the $n$-fold composition of the superoperator $\mathcal{L}: \mathcal{B}(\mathcal{H}) \to \mathcal{B}(\mathcal{H})$.

The form factor (\ref{dff-def}) is closely related to Uhlmann's ``transition probability"\cite{Uhlmann1976}, or mixed state fidelity \cite{Jozsa1994}  
\begin{align}
f(t) = \tr\left( \rho(t) \rho(0)\right),
\end{align}
which provides a measure of decoherence due to interaction with an environment \footnote{Recently dubbed the Uhlmann fidelity \cite{Tonielli2018}, but also referred to as the mixed-state memory fidelity \cite{Lidar1998,Bacon1999}, relative purity \cite{DelCampo2013}, linear fidelity, or simply the overlap \cite{Audenaert2012}.}. In the thermodynamic limit, where the system's Hilbert space dimension $N \to \infty$, the ensemble-averaged fidelity for an initial pure state is given by 
\begin{align}
\langle f(t) \rangle & = \frac{1}{N+1}\left[1 + N F(t) \right].\label{fidelity-avg}
\end{align}
The asymptotic behavior of Uhlmann's fidelity was considered in \cite{Tonielli2018} and shown to exhibit non-trivial scaling behavior tied to the non-orthogonality of eigenvectors of the Lindblad generator. As a result of the ensemble averaging, the dissipative form factor, and thus the fidelity, does not carry any direct information about eigenvectors. Our results for $F(t)$ can be imported into \eq{fidelity-avg} to recover the long-time behavior of the fidelity. In Refs. \cite{Miller1998a,Tameshtit1993}, \eq{fidelity-avg} was referred to as the survival probability function \cite{DelCampo2013}. The adjoint Lindblad generator $\mathcal{L}^{\dagger}$ (which is the conjugate transpose in the matrix representation) governs the time evolution of observables $A(t) = e^{ t\mathcal{L}^{\dagger} } A(0)$, and can be used to compute ensemble-averaged autocorrelation functions $C(t) = \langle A(t) A(0)\rangle$ (assuming $A(0)$ is traceless for simplicity)
\begin{align}
C(t) = \frac{N^{2}F(t) - 1}{N^{2}-1} C(0),
\end{align}
once again illustrating the central role of the DFF in the ensemble-averaged dynamics. Most significantly, such autocorrelation functions can be measured in NMR experiments on echo dynamics \cite{Goussev2012}. 

It is also useful to introduce here the {\it spectral gap} of the Lindblad superoperator:

\begin{defn}[Spectral Gap]
	For finite $N$, the spectral gap is defined as
	\begin{align}
\Delta =  - {\rm max}\,{\rm Re} ( z_{i}), \quad z_{i} \in \sigma(\mathcal{L})/\{0\},\label{gap-def}
\end{align} 
where $\sigma(\mathcal{L})/\{0\}$ denotes the eigenvalue spectrum of the Lindbladian excluding the zero modes.
\end{defn}

 Since ${\rm Re}(z_{i}) \le 0$ for all eigenvalues of $\mathcal{L}$, the spectral gap is a positive real number. At finite $N$ and for a quenched random Lindbladian, the spectral gap follows directly from the limit \footnote{This definition is sufficient for our purposes since the steady state is unique and purely imaginary eigenvalues require additional structure and symmetry, and thus have a vanishingly small probability to occur in our ensembles. } , 
 
 \begin{align}
 \Delta = - \lim_{t \to \infty} \, \frac{1}{t} \log \left[ \frac{1}{N^{2}} \left( \tr \, K_{t} - 1\right) \right].	
 \end{align}
 
 In Appendix (\ref{sec:replica}), we argue that in the large $N$ limit, the ensemble averaged spectral gap will be given by
 
 \begin{align}
 \langle \Delta \rangle = - \lim_{t \to \infty} \, \frac{1}{t} \log \, F(t) \, + O(N^{-1}). \label{gap-limit}
 \end{align}

This can be viewed as a reflection of the self-averaging property of the trace of the propagator. Consequently, we identify the {\it asymptotic decay rate}, defined by the right-hand-side of Eq.(\ref{gap-limit}) with the expected value of the spectral gap. Exponential relaxation of the DFF then implies a nonvanishing average spectral gap, whereas algebraic relaxation implies a vanishing gap.


We state our main results as propositions below. To get there, we must first introduce the concept of a non-crossing truncation of the Lindblad superoperator.

\begin{defn}[non-crossing truncation]\label{noncrossing}
	The non-crossing truncation of the simple dissipator is defined by
	\begin{align}
	\tilde{\mathcal{D}}_{L} = - L^{\dagger}L\otimes_{t} \mathbbm{1} - \mathbbm{1} \otimes_{t} L^{\dagger}L,
	\end{align}
which consists of removing the recycling term $2 L\otimes_{t} L^{\dagger}$. The non-crossing truncation of the Lindbladian $\tilde{\mathcal{L}}$ consists of replacing all simple dissipators with their non-crossing truncation.
\end{defn}

The terminology ``non-crossing" comes from the diagrammatic analysis we develop in Sec.(\ref{sec:gin-diss}), where its meaning is made clear. Briefly, the density matrix can be viewed as an element of $\mathcal{H}_{L}\otimes \mathcal{H}_{R}$, where $\mathcal{H}_{L}$ and $\mathcal{H}_{R}$ are left (bra) and right (ket) identical  copies of the Hilbert space. Some operators appearing in the Lindbladian act as the identity on one of these spaces, such as the term $L^{\dagger}L\otimes_{t} \mathbbm{1}$ which acts trivially on the right Hilbert space. On the other hand, the recycling term $L\otimes_{t} L^{\dagger}$ acts nontrivially on both the left and right Hilbert space. The non-crossing truncation involves just removing the terms in the Lindbladian which act nontrivially on {\it both} left and right Hilbert spaces; these are precisely the recycling terms.

Our next result is crucial for establishing most of the following propositions (unless otherwise noted):

\begin{lemma}\label{lemma}
Let the jump operators be complex Ginibre random matrices \cite{Ginibre1965}. To leading order in $N$, the trace moments of the Lindblad superoperator are determined by the non-crossing truncation
\begin{align}
M_{n} =\lim_{N \to \infty} \frac{1}{N^{2}} \left\langle {\rm tr} \, \tilde{\mathcal{L}}^{n} \right\rangle + O(N^{-1}).
\end{align}

\end{lemma}
Our proof of this lemma relies on the graphical calculus developed in Sec.(\ref{sec:gin-diss}). The proof is completed in two stages: first for a single jump operator in Sec.(\ref{sec:gin-diss}), and next for multiple jump operators in Sec.(\ref{sec:multi-diss}). These proofs are recapitulated using a topological expansion in Appendix (\ref{sec:top_expansion}). The proof of the lemma for a simple dissipator is a stepping stone to the first proposition:

\begin{prop}[Simple Dissipator: complex Ginibre ensemble]\label{Result1} Consider a simple generator \eq{dissipator} with the jump operator $L$ a complex Ginibre random matrix, i.e. $L_{ij} \in \mathbbm{C}$ are i.i.d. complex Gaussian variables with variance $\langle |L_{ij}|^{2}\rangle = v/N$. The dissipative form factor for this generator is
\begin{equation}\label{F-single-jump}
F(t) = \frac{1}{N^{2}}\left\langle {\rm tr} \, e^{ \mathcal{D}_{L} t} \right\rangle  = e^{ - 4 v t} \left( I_{0}(2v t) + I_{1}(2vt) \right)^{2} + O(N^{-1}),
\end{equation}

where $I_{n}(x)$ is the modified Bessel function of the $n^{th}$ kind.
\end{prop}

In particular, at long times, the dissipative form factor for the simple dissipator decays as a power-law $F(t) \sim t^{-1}$. Therefore, the spectral gap {\it closes} for a simple dissipator with complex Ginibre jump operators. We present numerical evidence that this closing occurs as $\Delta \sim N^{-2}$ in Sec.(\ref{sec:gin-diss}), implying a diffusive dynamical exponent for the relaxation time $\tau \sim N^{z}$ with $z = 2$ \cite{Gier2006}. It is also worth mentioning that $F(\infty) = N^{-2}$, as a consequence of the existence of a unique stationary state $\mathcal{D}_{L}[ \rho_{ss}] = 0$ \cite{Baumgartner2008}. 

A close relative of the complex Ginibre ensemble is the ensemble of random normal matrices \cite{Zabrodin2004}. Both ensembles have the same eigenvalue joint probability distribution function, but their effect on the phenomenology of the dissipator is quite distinct. While the dissipator with Ginibre jump operators has a unique steady state, there are exactly $N$ stationary states for normal matrices. The dissipative form factor also exhibits different power-law relaxation in the two cases, which we state presently.

\begin{prop}[Simple Dissipator: Random normal matrix]\label{Result1b} Consider a simple generator \eq{dissipator} with the jump operator $L$ a random normal matrix $[L,L^{\dagger}] = 0$. The exact dissipative form factor for finite $N$ is 
\begin{equation}
F(t)  = \frac{1}{N} + \frac{1}{t^{2}} \left( 1 - \left( 1 + \frac{t}{N}\right)^{-N}\right)^{2} - \frac{1}{t^{2}} \left( 1 -  \frac{(1 + 2 t/N)^{N}}{(1 + t/N)^{2N}} \right).	\label{ff-normal-exact}
\end{equation}

At $N = \infty$, the second term takes over and $F(t) = (1 - e^{ - t})^{2} / t^{2}$.
\end{prop}
The derivation of (\ref{Result1b}) appears in Appendix (\ref{normal}), and does not require Lemma (\ref{lemma}), using instead exact results for the eigenvalue density correlation functions in the random normal matrix ensemble \cite{Zabrodin2004}.

The only difference between the Ginibre and normal matrix ensembles is the eigenvector statistics, which is trivial for the latter. It is natural that non-orthogonality of eigenvectors in the Ginibre case leads to a slower rate relaxation. Furthermore, the relaxation time with normal jump operators scales as $\tau \sim \sqrt{N}$, which means the steady state is achieved much sooner than in the Ginibre case. 

For the sake of comparison, we also discuss the simple dissipator with GUE jump operators in Sec. (\ref{sec:diss-gue}). The spectral gap here also closes, and the form factor scales as $F(t) \sim t^{-1/2}$ at long times. Nevertheless, the relaxation time $\tau \sim N^{2}$, as in the Ginibre case. 

The simple dissipator turns out to be rather exceptional in its algebraic relaxation to equilibrium. As we see next, considering even a single additional jump operator in \eq{dissipator} immediately leads to exponential relaxation, indicative of the opening of a thermodynamic spectral gap.

\begin{prop}[Multiple Jump Operators]\label{Result2} Let $H= 0$ and $m$ be finite such that $m/N \to 0$ as $N \to \infty$. For $a = 1, ..., m$, let the jump operators $L_{a}$ be independent complex Ginibre matrices with identical variance as in Prop.(\ref{Result1}). The dissipative form factor in the long-time limit is given by 
\begin{equation}\label{F-multi-jumps}
F(t) = C t^{ -3} e^{ - \Delta v t}  + O(N^{-1}),
\end{equation}
for a constant $C = O(1)$ given in Eq.\eq{multi-diss-ff}. Here, the asymptotic decay rate $\Delta$ is determined by 
\begin{align}
\Delta = 	2 (1 - \sqrt{m})^{2}.
\end{align}
\end{prop}
The proof requires Lemma (\ref{lemma}) and appears in Sec.(\ref{sec:multi-diss}) (see also Appendix (\ref{sec:top_expansion})). The exponential relaxation leads to a nonzero average spectral gap, per Eq.(\ref{gap-limit}).

The existence of a unique steady state for the dissipator with multiple jump operators implies again the asymptotic limit $F(\infty) = N^{-2}$ \cite{Baumgartner2008a}. For finite systems, the exponential relaxation indicates that this expression is valid on relaxation time scales which are roughly $\tau \sim  \frac{2}{\Delta}\log N$.

Finally, we consider a simple Lindbladian which consists of a Hamiltonian and a single jump operator. 

\begin{prop}[Simple Lindbladian]\label{Result3} Let the Hamiltonian $H$ be drawn from the Gaussian unitary ensemble, i.e. $H_{ij} \in \mathbbm{C}$ are i.i.d. complex Gaussian variables such that $H_{ij} = \bar{H}_{ji}$, with zero mean and variance $\langle |H_{ij}|^{2} \rangle = 1/N$. Let $m = 1$ with $L$ a complex Ginibre matrix with variance $\langle |L_{ij}|^{2}\rangle = 1/N$. Consider the simple Lindbladian $\mathcal{L} = \mathcal{C}_{H} + \gamma \mathcal{D}_{L}$, where the first and second terms are defined in Eq. (\ref{von-Neumann}) and Eq. (\ref{dissipator}), respectively, with $\gamma \in \mathbbm{R}^{+}$ a positive real-valued constant independent of $N$.

The large $N$ form factor at long times is bounded above by
\begin{equation}\label{F-ham}
F(t) \le   \frac{1}{4\pi^{2}} t^{-4} e^{ - \Delta t}  + O(N^{-1}), \quad t\to \infty,
\end{equation}
with the asymptotic decay rate $\Delta = 2 \delta$, where $\delta$ is the spectral gap of the corresponding non-Hermitian Hamiltonian $H - i \gamma L^{\dagger}L$.
\end{prop}

 The inequality for the DFF implies a lower bound on the average spectral gap which is finite in the thermodynamic limit. That the random non-Hermitian Hamiltonian should have a spectral gap is a remarkable fact first discovered in \cite{Haake1992}. This proposition shows that the leading order behavior of the form factor is determined by the non-unitary dynamics induced by the effective non-Hermitian Hamiltonian.
 
Lemma (\ref{lemma}) is necessary to show that the asymptotic decay rate is determined by the non-Hermitian Hamiltonian, and the derivation of the inequality for the resulting form factor is given in Sec.(\ref{sec:ham-diss}). The functional form of this decay rate was found in \cite{Haake1992,Lehmann1995}, and is re-derived in Appendix (\ref{sec:nonhermitian-ham}) using straightforward algebraic techniques. In particular, the decay rate has the asymptotic scaling $\Delta \sim 2\gamma$ for small $\gamma$, and $\Delta \sim 2^{1/3}(\gamma)^{-1/3}$ at large dissipative coupling. We conjecture that the spectral gap for the simple Lindbladian will obey the same scaling behavior. The observation of exponential decay at large dissipative coupling is reminiscent of the quantum Zeno effect (see e.g. \cite{Mensky2013,Popkov2018}), though the non-analytic scaling of $\Delta(\gamma)$  is perhaps unusual in this context.

The form factor for the simple Lindbladian is in fact a damped oscillating function of time. In the $\gamma = 0$ limit, the unitary evolution produces a dynamical scaling $t^{-3}$, well known from the spectral form factor averaged over the GUE (see (\ref{sff-H})). In the opposite limit, Prop.(\ref{Result1}) shows that the power-law becomes  $t^{-1}$. 

In Appendix (\ref{sec:lind-m}) we comment briefly on how the asymptotic decay rate, and thus the mean spectral gap, is expected to change when there are $m$ jump operators and a Hamiltonian. In this case, we find $ \Delta \sim  2m \gamma$ for small $\gamma$, and $\Delta \sim 2\gamma (1- \sqrt{m})^{2}$  for large $\gamma$. 

Our methods do not grant access to the detailed structure of the eigenvalue distribution of the Lindblad superoperator. The reason is that the moment generating function (DFF) is the Laplace transform of the holomorphic Green's function, whose domain of validity is outside the support of the eigenvalue density. Nevertheless, the holomorphic Green's function is sufficient to extract one of the most basic properties of Lindblad evolution, which is its mixing behavior, and can thus identify the existence of a spectral gap.

\subsection{Discussion and Background}\label{sec:discussion}

Our main results support the claim that generic Lindblad evolution  relaxes exponentially, and is thus rapidly mixing. This behavior gives strong evidence for the existence of a spectral gap in the spectrum of the Lindblad superoperator. Exponential relaxation of the form factor in the limit of large $N$ implies that the bulk spectrum (sometimes called the essential spectrum) is separated from the imaginary axis by a finite gap. However, it does not exclude the possibility of a subextensive $o(N^{2})$ number of isolated eigenvalues residing within this bulk spectral gap (see \cite{Dellnitz2000} for examples of such isolated eigenvalues of the Frobenius-Perron operator in classical dynamics). Such isolated eigenvalues were indeed found to determine the asymptotic decay rate in a companion paper Ref.\cite{can}, which considers a simple Lindbladian with $H$ and $L$ drawn from the Gaussian orthogonal ensemble (GOE). In particular, we found that while the spectral gap is always finite in the thermodynamic limit, for large values of dissipative coupling the spectral gap is determined by a single isolated eigenvalue which appears to split off from the bulk spectrum. While preliminary investigations do not reveal such an isolated mode for the ensembles considered in this paper, its occurrence is not strictly ruled out.

On the other hand, algebraic relaxation of the dissipative form factor is a clear signature of the spectral gap {\it closing} in the thermodynamic limit. We have seen that this occurs for simple dissipators (Props.(\ref{Result1}) and (\ref{Result1b})), whether the jump operators are non-hermitian, normal, or hermitian, and have checked this for these three cases with full random matrices. The power-law is different for all three, scaling as $t^{-1}$ (Ginibre), $t^{-2}$ (normal) and $t^{-1/2}$ (GUE). These disparate relaxing behaviors lead to relaxation times of the form factor which scale as $\tau  \sim N^{z}$, with a diffusive dynamical exponent $z = 2$ for the Ginibre and GUE, and $z= 1/2$ for the normal jump operators. 

For the Hermitian jump operators, the relaxation time is in fact the Heisenberg time, since the mean spacing of eigenvalues of the dissipator is $N^{-2}$. This interpretation is not directly applicable to the dissipator with non-hermitian jump operators, since in this case the eigenvalues are complex valued. However, as we show later, the complex Ginibre relaxation is controlled by its ``non-crossing truncation", which consists of Wishart matrices. In this setting, the level spacing once again scales as $N^{-2}$. Such relaxation times are observed in closed quantum systems as well  \cite{Dymarsky2018a,Schiulaz2018}. 

As for the random normal matrix, the unusual scaling of the relaxation time suggests that the asymptotic relaxation rate is controlled by the boundary of the distribution of eigenvalues, along which the eigenvalue density (which is a proxy for level spacing) is expected to scale like $1/\sqrt{N}$.

Finally, we remark that the time scales we have access to in our asymptotic analysis are likely semi-classical in nature. Here we give some heuristic justification for this statement. In Ref.\cite{Casati1997}, it was argued that for open chaotic systems, the survival probability, which is known classically to exhibit exponential decay, experiences significant quantum fluctuations on an intermediate time scale that can be parametrically smaller than the Heisenberg time, but still diverging with $N$. The classical dynamics then holds for times up to this quantum time scale, which is set by the mean spacing between scattering resonances on the complex plane (as opposed to the energy level spacing which sets the Heisenberg time). In our asymptotic analysis, we work with the limit in which the mean spacing of the Lindblad eigenvalues tends to zero. Furthermore, the effective Hamiltonian describing the simple Lindbladian (i.e. $K$ in Eq.(\ref{eff-ham})) can be interpreted in the context of scattering theory, and describes $N$ bound states coupled via $N$ decay channels to the continuum. An extensive number of decay channels corresponds to a semi-classical limit in the chaotic scattering problem, and generically leads to exponential decay in e.g. bound state survival probability  \cite{Lewenkopf1991,Savin1997}. Finally, our results are qualitatively similar to those appearing in classical chaotic scattering \cite{Motter2002}. For these reasons, we speculate that large $N$ random Lindbladians display dynamics which is morally classical, despite not necessarily having a clear classical limit. This conclusion is also supported by the argument in Ref.\cite{xu2019extreme} that random dissipators (specifically with Hermitian jump operators) have a maximal decoherence rate (at short times), and thus should naturally lead to classical evolution at late times.

\smallskip
\paragraph*{Rapid Mixing and Spectral Gap in Open Quantum and Classical Systems}

Having summarized and discussed the main results of our paper, we now proceed to attempt giving our work meaning by properly fleshing out the context. There is a deep and extensive literature on related questions in the setting of classical and quantum dynamics that we now review. 

In the broadest context, our results concern the property of {\it mixing} in dynamical systems, which appears commonly in ergodic theory \cite{openbook2006}. The mixing time is the characteristic time scale required to reach the stationary state or equilibrium. For classical and quantum Markov processes, exponential relaxation toward steady state is referred to as {\it rapid mixing} \cite{Kastoryano2013,Lucia2015}. 

For a quantum Markov process, the rapid mixing property is controlled by the spectral gap (\ref{gap-def}) of the Lindblad generator, sometimes referred to as the {\it dissipative gap} \cite{Znidaric2015}. For finite-dimensional systems, the spectral gap  determines the slowest decay rate. In the thermodynamic limit, the spectral gap might close, leading to the possibility of observing algebraic relaxation of certain observables or correlations functions \cite{Medvedyeva2014,Cai2013}, and also potentially signaling a dissipative quantum phase transition \cite{Prosen2008,Kessler2012}. The role of the spectral gap in relaxation times of Lindblad dynamics was studied extensively for a particular class of boundary-driven Lindblad evolution in \cite{Znidaric2015}. 

The gap also has implications for the stability of the resulting steady state to perturbations of the generator of time evolution \cite{Cubitt2015,Lucia2015}, as well as for the clustering of correlations between local observables \cite{Poulin2010,Kastoryano2014}. In Ref.\cite{Temme2013}, the existence of a spectral gap is proven for a certain class of Davies generators, which are generators of Lindblad form constructed explicitly to relax to thermal equilibrium. 

Random quantum channels are also known to exhibit a spectral gap. In this setting, the steady state is an invariant state under the map, and has eigenvalue equal to unity. The rest of the eigenvalues are contained in the unit disk. If the spectral radius (i.e. largest eigenvalue in absolute value) is strictly less than unity, these operators are said to have a spectral gap. Using the Kraus representation, Ref.\cite{Bruzda2008,Bruzda2010} proposed a random ensemble of quantum channels based on their action on the orthonormal basis of $SU(N)$ generators. The result for a quantum channel with $m$ Kraus operators was a spectrum resembling the real Ginibre matrix ensemble \cite{Ginibre1965}, supported on a circle with radius $ 1/\sqrt{m}<1$.

Random quantum channels induced by Haar-distributed random isometries were introduced in \cite{Hayden2008} and studied extensively in the quantum information context \cite{Collins2015}. Recently in Ref.\cite{Gonzalez-Guillen2018}, such maps were shown to have a spectral gap in the limit of large Hilbert space dimension and for a fixed number of Kraus operators. This follows on previous work \cite{Hastings2007} which proves a spectral gap for a quantum channel with a fixed number of Haar distributed unitaries.

A similar approach considers maps describing open quantum systems obtained by projecting an $N$ dimensional random unitary onto a smaller $M$ dimensional subspace, resulting in a ``truncated" unitary matrix. The spectrum of truncated random unitaries was studied in \cite{Zyczkowski2008}, and also shown to exhibit a spectral gap in the $N\to \infty$ limit when $M/N$ is kept fixed. 

The spectral gap appears to be a manifestation of quantum chaos, whereas its closing indicates the emergence of integrability. In Ref.\cite{Novaes2013}, the projection technique was used to quantize a classical open chaotic map. They observed a spectral gap in the quantum map, indicating a minimal resonance lifetime. Ref.\cite{Nonnenmacher2011} determines a condition for the existence of a spectral gap in quantized open chaotic maps (i.e. quantum maps which have a clear chaotic classical limit) related to a quantity known as the topological pressure, characterizing the dynamics that is recovered in the classical limit \cite{Gaspard1989}. 

 In the classical setting, the dynamical map acting on the probability density function is the Frobenius-Perron (FP) operator, and the eigenvalues are known as the Pollicott-Ruelle (PR) resonances. Refs. \cite{Nonnenmacher2003} and \cite{Saraceno2004} consider quantum maps with well-defined classical limits, and found that the PR resonances governed the asymptotic decay rates of the quantum system. In \cite{Nonnenmacher2003}, it was argued that a spectral gap in the quantum map is symptomatic of ergodicity and mixing of its classical limit, whereas a closing spectral gap appeared to coincide with integrability. Classical chaotic scattering also exhibits a gap in the distribution of resonances, related to the topological pressure, also observed experimentally \cite{Barkhofen2013}.

Indeed, the dynamical consequences of proximity to integrability have been studied in the classical setting. Ref.\cite{Vivaldi1983} observe a crossover from exponential to power-law decay of auto-correlation functions in the stadium billiard, connected to the presence of arbitrarily long regular motion in the evolution of stochastic (ergodic) orbits. 

Classical chaotic scattering is also known to show a transition from exponential to power law decay of survival probability of a particle in a scattering region, depending on the existence of KAM tori in phase space (see e.g. \cite{Barra2002} or the monograph \cite{Gaspard2005}). Our results are curiously consonant with the conclusions of Ref. \cite{Motter2002} on classical chaotic scattering, which argues that algebraic decay is unstable to the presence of dissipation, and exponential decay takes over for {\it any} amount of dissipation. In the context of open quantum systems, we find that Hamiltonian evolution with the addition of dissipation leads to exponential decay of fidelity (\ref{fidelity-avg}), which becomes the survival probability for pure Hamiltonian evolution. These results suggest that random Lindblad dynamics in the large $N$ limit are somehow semi-classical in nature, and capture qualitative features of fully chaotic classical systems.

There do exist some noteworthy connections between dissipative chaotic systems and random matrix theory. In dissipative quantum systems with a chaotic classical limit, the eigenvalues of the dynamical map were argued to exhibit a universal cubic level repulsion \cite{Grobe1989}, reviewed in the book \cite{Haake2001} (see also \cite{Braun2001}). A semiclassical expression involving periodic orbits was found for the trace of a discrete time dynamical map for the dissipative kicked rotor (analogous to our dissipative form factor) in  \cite{Braun1999},  (for a nice review, see the monograph \cite{Braun2001}). The spectral gap in the dissipative kicked rotor model was shown to be determined by the leading Pollicot-Ruelle resonance of the FP operator \cite{Braun2001,Braun1999b}.

Quantum dynamics in the presence of random environments or interactions has also been studied in the past \cite{ko1976one,lebowitz2004random}. Notably, in \cite{Gorin2004} the fidelity is studied in the linear response regime assuming a random matrix perturbation of the time-reversed Hamiltonian. Their result shows the universal relaxation behavior of quantum chaotic systems. It would be interesting to connect these results to the mixed state fidelity (\ref{fidelity-avg}) for random Lindblad evolution.

Finally, we mention a conjecture for the Lindbladian spectral gap motivated by the comparison to chaotic scattering theory. It is known that for a small number of decay channels (non-extensive in Hilbert space dimension), the survival probability asymptotically decays algebraically in time \cite{Lewenkopf1991}. Translating to the Lindblad case, this corresponds to the setting in which the jump operators are low-rank matrices. At present, we are not able to extend our diagrammatic expansion to low-rank random matrices, but we can make the following conjecture: the spectral gap of the simple Lindbladian will {\it close} in the large $N$ limit for jump operators with a fixed rank that does not grow with $N$.

\section{Simple Dissipator with a Random Jump Operator}\label{sec:dissipator}

\subsection{Complex Ginibre Ensemble}\label{sec:gin-diss}
In this section we provide part of the proof of Lemma (\ref{lemma}), and as a first application prove Prop. (\ref{Result1}). The techniques which are developed in this section will be called upon later to provide the proofs of Props. (\ref{Result2}) and (\ref{Result3}). Our approach to studying the dissipative form factor is to develop a diagrammatic calculus for calculating moments of the dissipator (for now dropping the subscript appearing in \eq{dissipator} for notational convenience)
\begin{align}
\mathcal{D} = 2 L \otimes_{t} L^{\dagger} - L^{\dagger} L \otimes_{t} \mathbbm{1} - \mathbbm{1} \otimes_{t} L^{\dagger}L,	\label{diss-tensor}
\end{align}
when the jump operator is a complex Ginibre random matrix with i.i.d. complex Gaussian entries with zero mean and variance
\begin{align}
\langle \bar{L}_{ij} L_{kl} \rangle = \frac{v}{N} \delta_{ik}\delta_{jl}	, \quad L_{ij} \in \mathbbm{C}.\label{var-Gin}
\end{align}
We begin with some descriptive numerical results to help us anticipate the structure of our final result. The eigenvalues are entirely contained in the left-hand complex plane, and form a droplet which has maximal density close to the origin (see Fig.(\ref{fig:dissevals})). The hermiticity preserving property of the QDS generator $(\mathcal{L}[\rho])^{\dagger} = \mathcal{L} [ \rho^{\dagger}]$ also implies that the eigenvalues are either real or come in complex conjugate pairs. For this reason, the density is symmetric under reflection across the real axis. The distribution of eigenvalues is highly reminiscent of that for real Markov generators \cite{Timm2009, Bordenave2014}, which are real Ginibre matrices whose columns sum to zero \footnote{This similarity becomes essentially exact for a dissipator with $N^{2}-1$ jump operators, as shown recently in \cite{Denisov2018}}.

\begin{figure}[h!]
\centering
\includegraphics[scale=0.7]{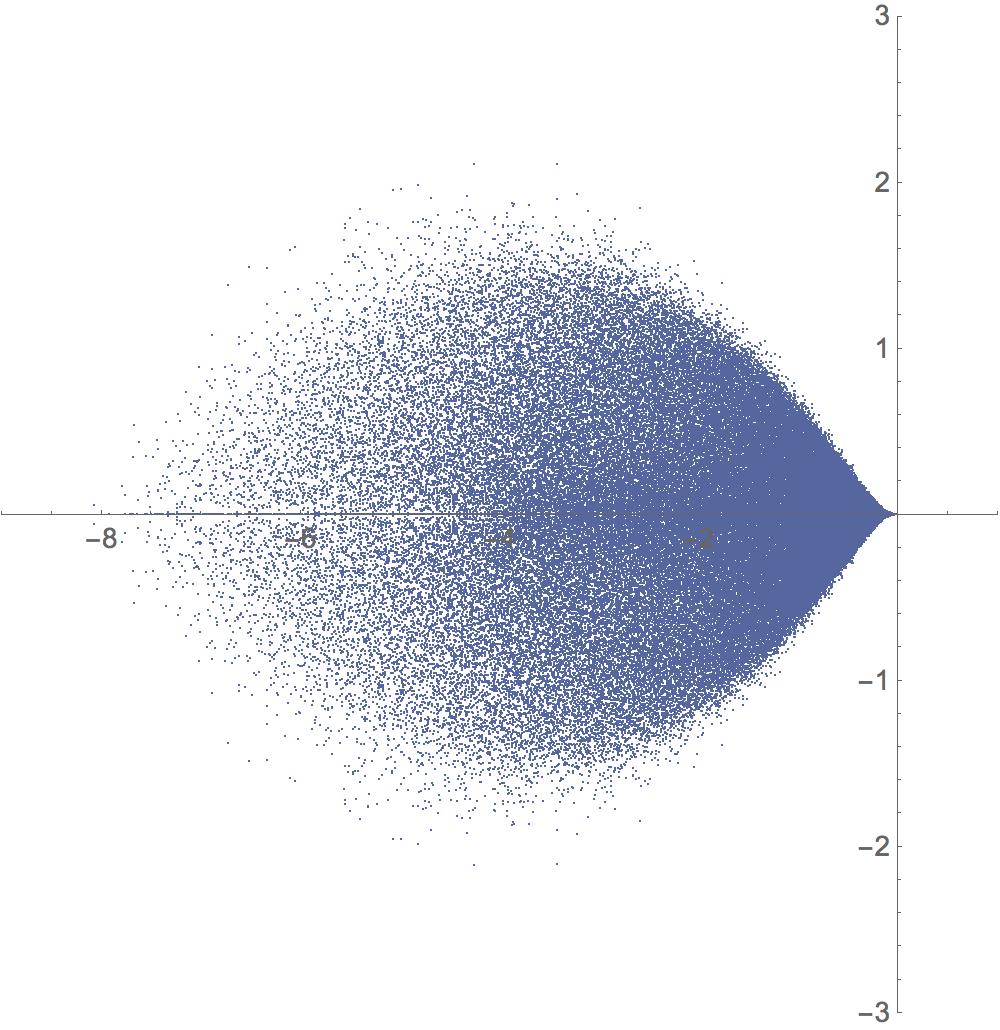}\\
\caption{\label{diss-gin-evals} Eigenvalues of simple dissipator with complex Ginibre jump operators with $v = 1$, $N  = 70$ for $25$ samples. }
\la{fig:dissevals}
\end{figure}

\begin{figure}
\centering
\includegraphics[scale=0.7]{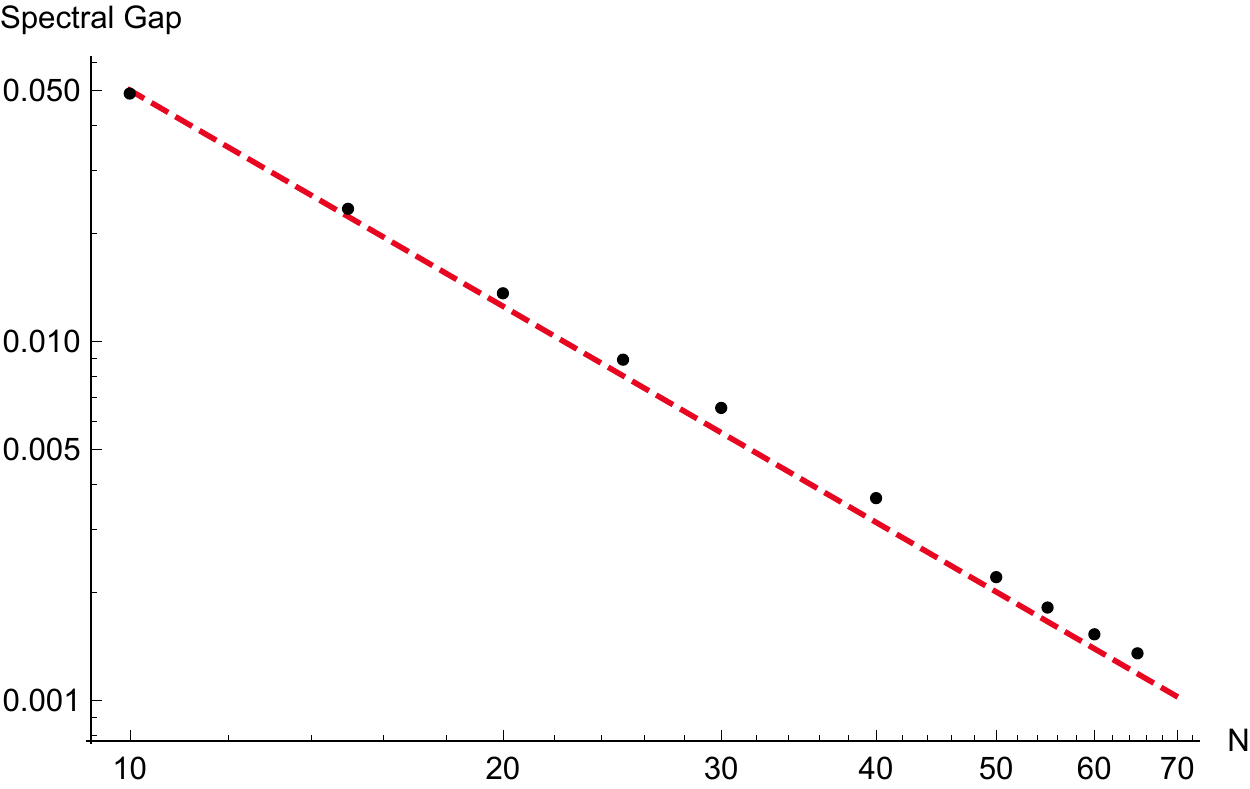}\\
\caption{Average spectral gap (black points) of \eq{diss-tensor} as a function of $N$ on log plot consistent with a closing of the spectral gap as $a N^{-2}$ (dashed red), shown with the empirical least-square fit value of $a \approx 5.016$.}
\label{fig:gap-gin-diss}
\end{figure}

The spectral gap exhibits a clear $N^{-2}$ scaling with Hilbert space dimension in Fig.(\ref{fig:gap-gin-diss}), indicating that in the thermodynamic limit the spectral gap will close. We prove this assertion below. 

It is not possible to describe the spectrum of (\ref{diss-tensor}) in a simple way in terms of eigenvalues or singular values of $L$, and we have not found an expression for the joint probability distribution function of its eigenvalues. Such shortcomings rule out the possibility of pursuing many classic approaches in RMT such as the method of orthogonal polynomials or the Coulomb plasma mapping. Instead, we appeal to another popular approach which involves developing a systematic large $N$ diagrammatic expansion of the moments. 

The method of moments was first used by Wigner to derive the semi-circle distribution of eigenvalues for full random Hermitian matrices \cite{Wigner1955}, and later for k-body random matrix ensembles \cite{pandey1979binary,mon1975statistical}. The large $N$ diagrammatics for computing moments goes back to Refs. \cite{t1974planar,brezin1978planar}, which develop a topological expansion capturing the $N$ scaling of graphs by their Euler characteristic. Similar techniques arise in considering moments of more structured operators, such as k-body random matrices \cite{mon1975statistical}, and propagators for systems with spatial locality constraints (for some recent examples see  \cite{chan2018solution,kukuljan2016corner}). Our ensembles of random Lindbladians with fullly random jump operators and Hamiltonians bear similarities to, but have important differences from, these previously studied models. While the planar approximation for traditional matrix models does not quite seem to hold due to the tensor product structure of Lindbladians, the lack of additional structure such as spatial locality allows for a simple revision of the classic topological expansion.

To begin, we discuss some basic symmetry properties of the moments. Expressed as a fourth-order tensor, the dissipator has the property
\begin{align}
\sum_{i=1}^{N}\mathcal{D}_{ii kl}	 = 0.
\end{align}
This is a consequence of the dissipator being trace-preserving, and it can be seen directly from the representation

\begin{align}
\mathcal{D}_{ijkl} = 2 L_{ik} L_{lj}^{\dagger} - ( L^{\dagger}L)_{ik}\delta_{jl} - ( L^{\dagger}L)_{lj}\delta_{ik}.
\end{align}
Furthermore, this property holds for all matrix moments of the dissipator

\begin{align}
\sum_{i}\left( \mathcal{D}^{n} \right)_{iikl} = 0,	\label{TP-moment}
\end{align}
where the meaning of $\mathcal{D}^{n}$ is the $n$-fold composition of the map with itself. In the tensor notation, for instance, 

\begin{align}
\left(\mathcal{D}^{2}\right)_{ijkl}	= \mathcal{D}_{ij ab} \mathcal{D}_{ab kl},
\end{align}
where summation over repeated indices here is implied. As a consequence of this symmetry, the propagator $K_{t} = e^{ t\mathcal{D} }$ satisfies

\begin{align}
\sum_{i}\left(K_{t}\right)_{iikl}  = \delta_{kl},
\end{align}
which is another statement of the conservation of probability, since it implies $\tr(\rho(t)) = \tr (\rho(0))$. This constraint will be preserved upon ensemble-averaging over the jump operators. The most general form for the average matrix moment of the dissipator is 
\begin{align}
\left\langle \left(\mathcal{D}^{n}\right)_{ijkl}	\right\rangle  = A_{n} \delta_{ik}\delta_{lj} + B_{n} \delta_{ij}\delta_{lj} + C_{n} \delta_{il}\delta_{jk}.
\end{align}
The structure of the matrix moment will necessarily be a sum of products of Kronecker delta functions, due to Wick contraction of jump operators with the free indices $\{i,j,k,l\}$. The coefficients $A_{n}$, $B_{n}$, are $C_{n}$ are then combinatorial factors which depend non-trivially on $n$ and $N$, and are proportional to $v^{n}$.

First, we will show that the coefficient $C_{n} = 0$. The matrix moment is the sum of many terms, all of the form
\begin{align}
X_{ik} Y_{lj} \label{gen-moment},
\end{align}
where $X$ is a string of $L$ and $L^{\dagger}L$, and $Y$ is a string of $L^{\dagger}$ and $L^{\dagger}L$. Ensemble averaging over the complex Gaussian matrices $L$ requires Wick contraction between pairs $L$ and $L^{\dagger}$. Therefore, in order for a contraction to produce $\delta_{il}$, we need terms of the form $L_{iq} L_{p l}^{\dagger}$, or conversely $L_{qi} L_{lp}^{\dagger}$. However, the structure of the moments implies that these combinations are impossible. From \eq{gen-moment}, all we can hope to find are combinations $L_{iq} L^{\dagger}_{lp}$ or $L_{iq}^{\dagger}L_{lp}$. Therefore, it is impossible in general to find a contraction producing $\delta_{il}$ (an identical argument applies to $\delta_{kj}$) and we must have $C_{n} = 0$. Note that these arguments fail for GOE jump operators, for which $C_{n}$ does not vanish. 

Next, the constraint \eq{TP-moment} implies that
\begin{align}
\sum_{i}\left\langle \mathcal{D}^{n}_{iikl}	\right\rangle  = A_{n} \delta_{ik}\delta_{li} + B_{n} \delta_{ii}\delta_{lj}  = \left(A_{n} + N B_{n}\right)\delta_{kl} = 0,
\end{align}
which is only possible if $A_{n} + N B_{n} = 0$. Using this, we see that the averaged trace moments of the dissipator are given by $A_{n}$,
\begin{align}
\frac{1}{N^{2} -1} \left\langle \tr \, \mathcal{D}^{n} \right\rangle	 = \left\langle \sum_{i,j} \, \mathcal{D}^{n}_{ijij} \right\rangle  = A_{n}.
\end{align}

The factor $N^{2}-1$ which appears in the denominator is the maximal number of non-vanishing eigenvalues of the dissipator, since there is a single unique zero mode \cite{Baumgartner2008}. A simple counting implies that for a variance which scales like $N^{-1}$ as in Eq.\eq{var-Gin}, the moment $A_{n}$ should be order $O(1)$ with $N^{-1}$ corrections. Thus, we find the general formula for the matrix moment in terms of the trace-moments

\begin{align}
\left\langle \mathcal{D}^{n}_{ijkl} \right\rangle 	 = A_{n} \left( \delta_{ik}\delta_{lj} - \frac{1}{N} \delta_{ij}\delta_{kl}\right).	
\end{align}

As a consequence, the average propagator has the power series expansion

\begin{align}
\left\langle K_{t}\right\rangle  &= \delta_{ik}\delta_{jl} + A_{1} t 	\left( \delta_{ik}\delta_{lj} - \frac{1}{N} \delta_{ij}\delta_{kl}\right)	 + \frac{A_{2}t^{2}}{2!} \left( \delta_{ik}\delta_{lj} - \frac{1}{N} \delta_{ij}\delta_{kl}\right)	+ ...\\
& = \frac{1}{N}\delta_{ij}\delta_{kl} + G(t) \left( \delta_{ik}\delta_{lj} - \frac{1}{N} \delta_{ij}\delta_{kl}\right),
\end{align}

where
\begin{align}
G(t) = \sum_{k = 0}^{\infty} \frac{A_{k} t^{k}}{k!},
\end{align}
is the moment generating function. Accordingly, the disorder-averaged density matrix obeys

\begin{align}
\langle \rho(t) \rangle = \frac{1}{N} \mathbbm{1} + G(t) \left( \rho(0) - \frac{1}{N} \mathbbm{1}\right).
\end{align}

This expression is somewhat misleading, since it seems to imply that the steady state is the uniform state proportional to the identity. However, it is known that the steady state is given by \cite{Baumgartner2008}

\begin{align}
\rho_{ss} = \frac{ M^{-1}}{\tr M^{-1}}, \quad M = L^{\dagger}L,
\end{align}
which is clearly not the uniform state. The discrepancy clearly comes from the order in which we do things. If we take an ensemble average before sending $t \to \infty$, then indeed the evolution tends to take the state to the infinite temperature, maximal entropy state. However, if we quench ``disorder" and run $t\to \infty$, the nature of the steady state is far from uniform, and in fact appears to have relatively low entropy (since $M$ tends to have small eigenvalues, the steady-state density matrix will generically have a single large eigenvalue). 

With the propagator in hand, we can calculate the dissipative form factor by taking the trace $\sum_{i,j} K_{t,ij ij}$ to find
\begin{align}
F(t) = \frac{1}{N^{2}} + (1 - \frac{1}{N^{2}}) G(t).	
\end{align}
Developing the generating function as an asymptotic series in $N^{-1}$, $G(t) = G_{0}(t) + N^{-1} G_{1}(t) + ...$, we have in the large $N$ limit
\begin{align}
F(t) =  G_{0}(t) + O(N^{-1}) .\label{F-expansion}
\end{align}

Our primary concern will be to understand the time evolution of $G_{0}(t)$. If this exhibits power-law decay, then we can say confidently that the eigenvalue spectrum of the dissipator is gapless. Having isolated the main object of study, we develop now the graphical calculus which is used to calculate the moments, and thus the generating function, to leading order. 

We begin with a simple example - the mean-field dissipator. It can be evaluated exactly to yield
\begin{align}
\langle \mathcal{D}_{ijkl}\rangle 	 = - 2  v \left( \delta_{ik}\delta_{jl} - \frac{1}{N}\delta_{ij}\delta_{kl}\right).\label{diss-first-order}
\end{align}

\begin{figure}[htbp!]
\centering
\includegraphics[scale=0.4]{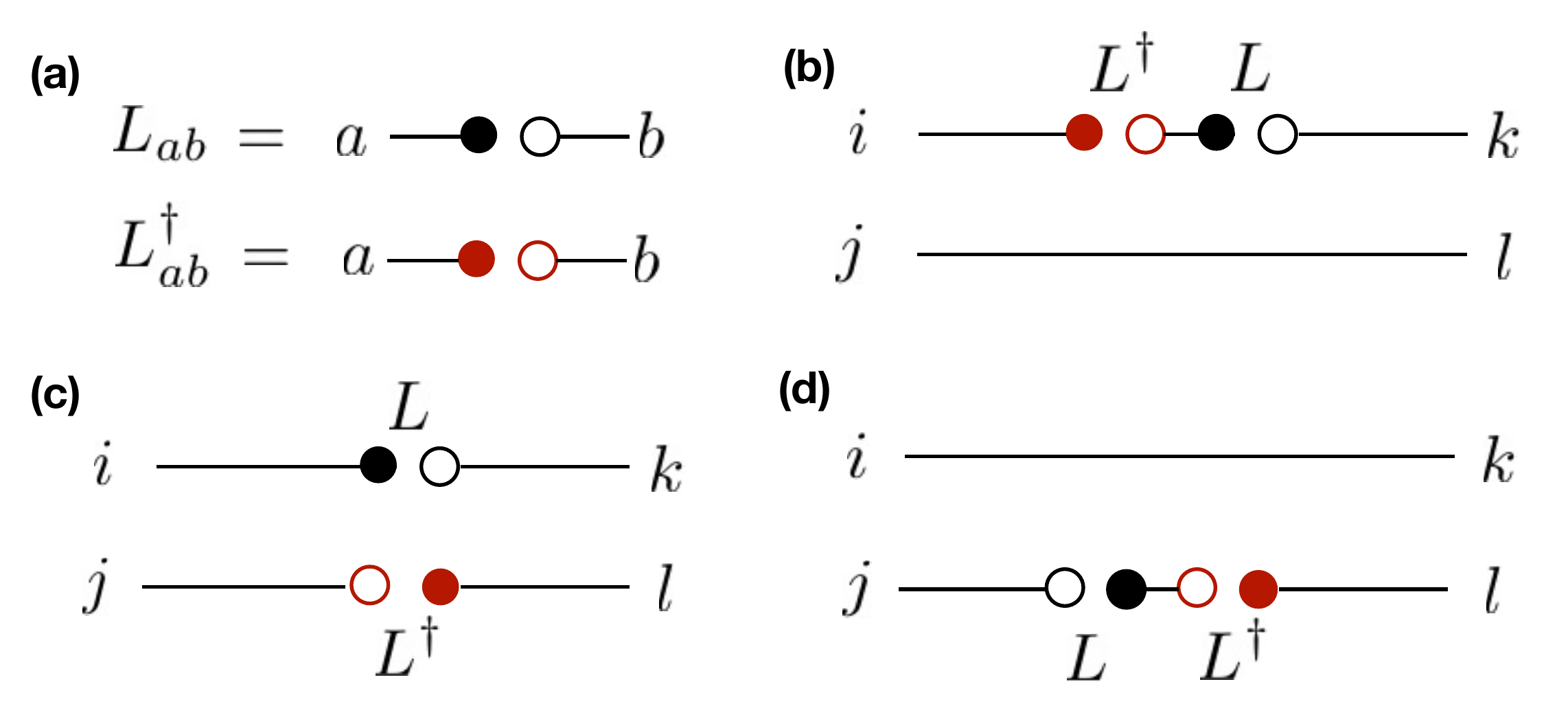}\\
\caption{\label{fig:diagrams-dissipator}The diagrams representing the dissipator have two legs. In a) we color $L$ (black) and $L^{\dagger}$ (red), and since they are not symmetric we differentiate the first and second index by filled and open dots. There are three diagrams  which are required to represent $\mathcal{D}_{ijkl}$ b) shows $(L^{\dagger}L)_{ik}\delta_{jl}$, c) shows $L_{ik} L_{jl}^{\dagger}$, and d) shows $\delta_{ik} (L^{\dagger}L)_{jl}$.  }
\end{figure}

The mean-field dissipator is the generator of a depolarizing channel \cite{Kastoryano2013},
\begin{align}
\langle \mathcal{D} \rangle \rho = - 2 v \left( \rho - \frac{1}{N} \tr (\rho )	\mathbbm{1}\right),
\end{align}
which has a spectral gap, and is thus rapidly mixing. We now show that taking fluctuations (higher moments) correctly into account will collapse the spectral gap. In order to do this, we introduce our  diagrammatic thinking. First of all, since we are dealing with fourth-order tensors, we need four external legs. The tensor product structure of the dissipator in (\ref{diss-tensor}) allows us to organize these four external legs into two edges. The identity matrix in this notation is $\delta_{ik}\delta_{jl}$ and is represented by two parallel lines connecting $i$ to $k$ and $j$ to $l$. From this starting point, we indicate the insertion of a jump operator by a pair of colored dots, as in Fig. (\ref{fig:diagrams-dissipator}a), representing the indices of the matrix $L$. Then we have $L^{\dagger}L\otimes_{t} \mathbbm{1}$ represented in Fig. (\ref{fig:diagrams-dissipator}b); $L\otimes_{t} L^{\dagger}$ in Fig. (\ref{fig:diagrams-dissipator}c); and $\mathbbm{1}\otimes_{t} L^{\dagger}L$ represented in Fig. (\ref{fig:diagrams-dissipator}d).

In taking the expectation value, we must perform a Wick contraction to evaluate the correlators of complex Gaussian fields. This is accomplished by connecting red to black, and filled to open. Figure ({\ref{fig:diagrams-first-order}) shows how (\ref{diss-first-order}) would arise diagrammatically.

\begin{figure}[htbp!]
\centering
\includegraphics[scale=0.4]{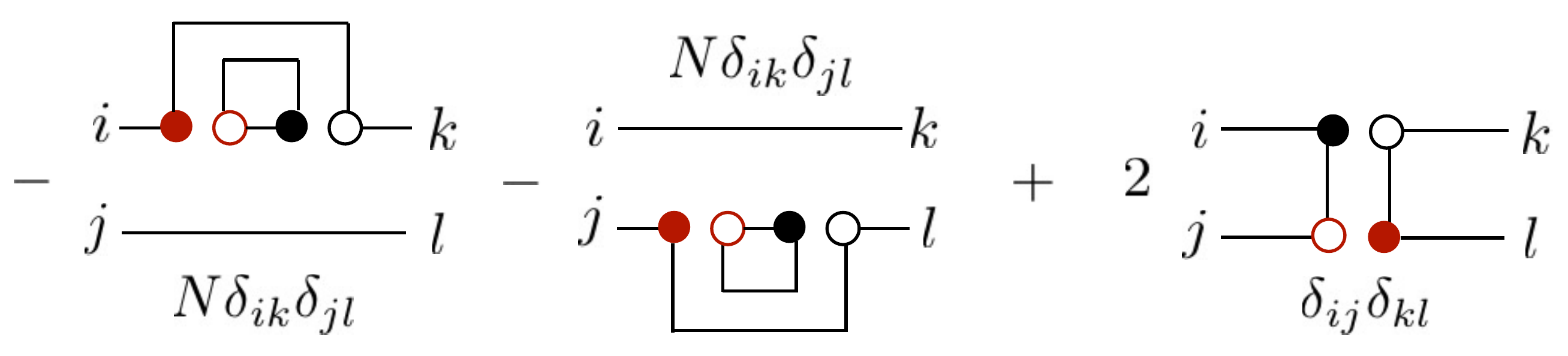}\\
\caption{\label{fig:diagrams-first-order}Diagrams for $\langle \mathcal{D}_{ijkl}\rangle$. The sum of diagrams must be multiplied by the variance $v/N$.}
\end{figure}

Equipped with these diagrammatics, we can compare the exact second moment of the dissipator with its diagrammatic representation. A direct, if tedious, calculation reveals that the second matrix moment is

\begin{align}
\left\langle \left(\mathcal{D}^{2}\right)_{ijkl}\right\rangle =  6 v^{2} \left( \delta_{ik}\delta_{jl} - \frac{1}{N} \delta_{ij} \delta_{kl}\right).	\label{second-moment-diss}
\end{align}

\begin{figure}
\centering
\includegraphics[scale=0.6]{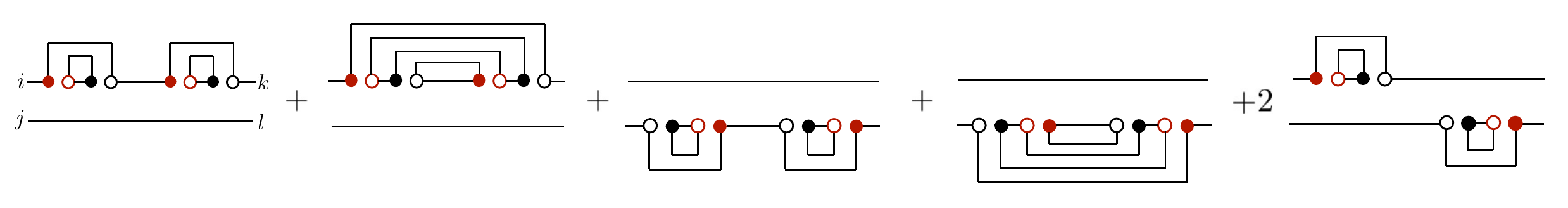}\\
\caption{\label{leading-diagram-dissipator} Leading order diagrams for the second moment with a single complex Ginibre jump operator. All diagrams are multiplied by the square of the variance $(v/N)^{2}$.}
\la{fig:leading-diagrams-second-order-2}
\end{figure}

The $O(1)$ second-moment diagrams are shown in Fig.(\ref{fig:leading-diagrams-second-order-2}) (all second-order diagrams are presented in Fig.(\ref{fig:diagrams-second-order-complete})). Since these must all be multiplied by the variance squared $(v/N)^{2}$, the infinite $N$ limit will preserve only diagrams of order $N^{2}$. This is only possible if under Wick contraction, the diagram includes two closed loops. It is clear that order $N^{2}$ terms in these second-order diagrams are only possible when the contractions occur on either the top or bottom edge, and not across the edges. Furthermore, such separated diagrams are not possible if there is any insertion of the recycling term $L\otimes L^{\dagger}$, for the simple reason that it only ever introduces an odd number of contractible Gaussian variables on a single edge, requiring a contraction that joins the edges. 

Having found the structure of the leading order diagrams, we are left with the challenge of correctly counting them. This is accomplished if we simply remove the recycling term when calculating the moment. Note that for the second moment, we can recover the leading order contribution by looking at the simpler problem  
\begin{align}
 \langle \tr \, \mathcal{D}^{2}\rangle = \lim_{N \to \infty}  \langle (L^{\dagger}L)^{2}\otimes_{t} \mathbbm{1} + L^{\dagger}L \otimes_{t} L^{\dagger} L + \mathbbm{1}\otimes_{t} (L^{\dagger}L)^{2}\rangle = \left\langle \left(L^{\dagger}L \otimes_{t} \mathbbm{1} + \mathbbm{1} \otimes_{t} L^{\dagger}L \right)^{2} \right\rangle .
\end{align}
More generally, these considerations lead us to conclude that  the leading order diagrams are completely accounted for by the ``non-crossing trunctation" of the dissipator 

\begin{align}
M_{n} =  \lim_{N\to \infty}\frac{(-1)^{n} }{N^{2}}\left\langle \tr \left( L^{\dagger}L\otimes_{t} \mathbbm{1} + L^{\dagger}L \otimes_{t} \mathbbm{1}\right)^{n}\right\rangle.\label{non-cross-body}
\end{align}

This concludes the proof of a special case of Lemma (\ref{lemma}) for simple dissipators. We finish the proof in Sec.(\ref{sec:multi-diss}) for multiple jump operators (see also Appendix (\ref{sec:top_expansion}}) for a more topologically oriented proof). Note that while our diagrammatic picture is exact, it does not reduce in any simple way to the familiar self-consistent Born approximation, an observation also made in Ref.\cite{Oganesyan2003} in attempting such an approach to calculate the self-energy for the Liouville-von Neumann operator. The culprit in their case and ours is clearly the double-legged structure of the ``bare propagator", which spoils the traditional topological expansion in terms of planar diagrams \cite{t1974planar,brezin1978planar,bessis1980quantum}. For more on the topological nature of the large $N$ limit see Appendix (\ref{sec:top_expansion}). Now, we proceed to evaluate the leading order form factor.

The non-crossing truncation in (\ref{non-cross-body}) significantly simplifies the problem, since now we can use the eigenvalues $u_{i}$ of $L^{\dagger}L$ (i.e. the squared singular values of $L$) to compute

\begin{align}
M_{n}= \lim_{N\to \infty} \frac{(-1)^{n}}{N^{2}} \left\langle \sum_{i,j}( u_{i} + u_{j})^{n} \right\rangle.
\end{align}
The distribution of these eigenvalues follows the Mar\v{c}enko-Pastur law for complex Wishart matrices 
\begin{align}
\rho_{w}(x) = \frac{1}{2\pi v }\sqrt{\frac{4 v}{x} - 1}, \quad x\in (0, 4v].\label{density-MP}
\end{align}
where we have $\int \rho_{w}(x) = 1$. We also note here that this distribution follows from the resolvent 

\begin{align}
G_{w}(z) &= \frac{1}{N}\left\langle \tr  \frac{1}{z - L^{\dagger} L} \right\rangle  = \frac{1}{2v} - \frac{\sqrt{ z^{2} - 4 v z}}{2 v z},\label{resolvent-MP}
\end{align}
which we will make use of later on. The leading order generating function then follows immediately, and is given by
\begin{align}
G_{0}(t)&= \lim_{N\to \infty}\frac{1}{N^{2}} \left\langle \sum_{i,j} e^{ - u_{i} t - u_{j} t} \right\rangle =  \int e^{ - (x + y) t} \rho_{w}(x) \rho_{w}(y) dx dy,\nonumber\\
&	 = e^{ - 4 v t} \left( I_{0}( 2v t) + I_{1}( 2v t) \right)^{2} .\label{ff-gin-leading}
\end{align}
Along with \eq{F-expansion}, this proves our first proposition (\ref{Result1}). The asymptotic behavior of the modified Bessel function implies that for long time, the exponential factor will disappear and the dynamics follows a power-law $F(t) \sim t^{-1}$. Since there is a unique steady state, the asymptotic limit of the form factor is $F(\infty) = N^{-2}$. Therefore, we expect the large $N$ behavior to be relevant on time scales $t^{-1} >> N^{-2}$, i.e. $t <<N^{2}$. However, finite size numerics indicates that a second regime takes over at times $t \sim O(N)$ interpolating between large $N$ behavior \eq{ff-gin-leading} and the asymptotic limit $N^{-2}$. This is an indication that the sub-leading $O(N^{-1})$ correction to $F(t)$.

\begin{figure}[htbp]
\centering
\includegraphics[scale=0.7]{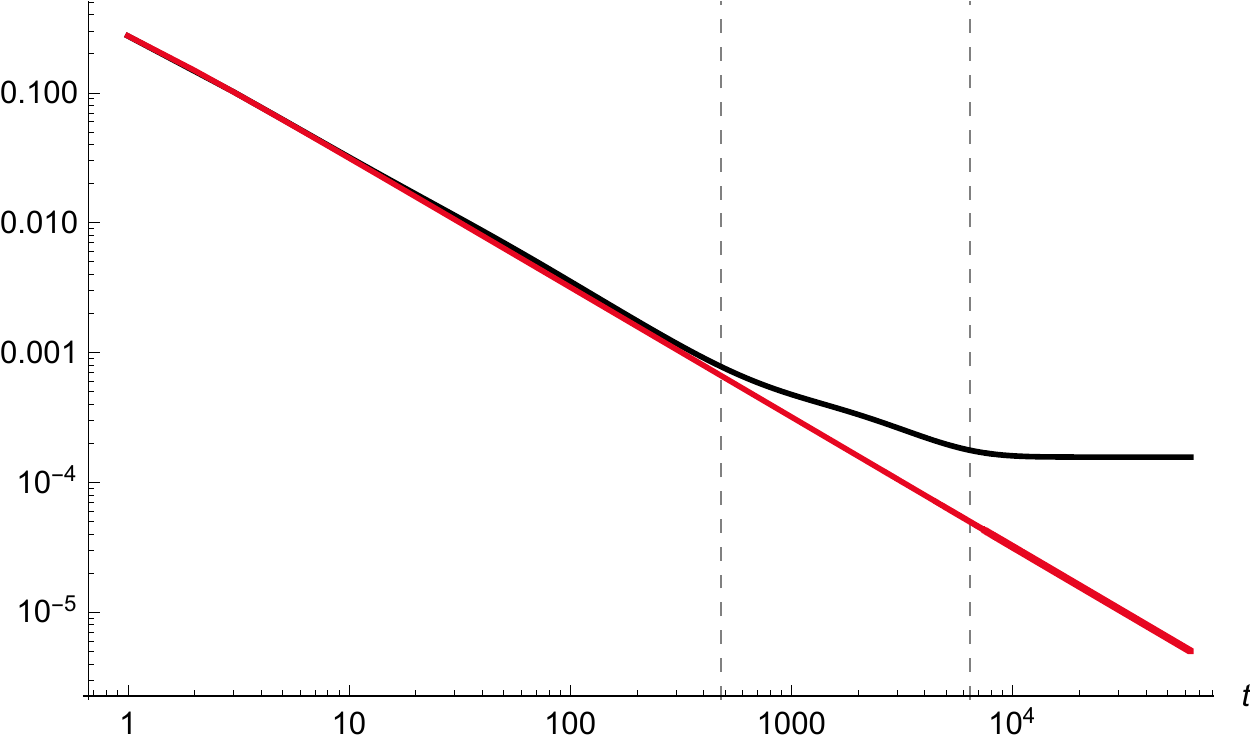}\\
\caption{\label{ff-ginibre-evals} Numerically exact dissipative form factor $F(t)$ (black) and the leading order asymptotic formula \eq{ff-gin-leading} (red) on a log scale for $N = 80$, $v = 1$, and a single sample. The dashed vertical bars, intended to guide the eye, appear at $6N$ and $N^{2}$. }
\la{fig:ffginibre}
\end{figure}

\subsection{Complex Random Normal Matrices}\label{sec:normal-jumps}

Here we consider a simple generator with the jump operator taken to be a random normal matrix satisfying $[L,L^{\dagger}] = 0$. The eigenvectors of a general normal matrix form an orthonormal basis $\langle \tau_{i}|\tau_{j}\rangle = \delta_{ij}$ where $L|\tau_{i}\rangle = \tau_{i} | \tau_{i}\rangle$. The eigenmodes of the dissipator are then $|\tau_{i}\rangle \langle \tau_{j}|$ for all $i,j = 1, ..., N$, with eigenvalues
\begin{align}
z_{ij} = 	 - |\tau_{i}|^{2} - |\tau_{j}|^{2} + 2 \tau_{i} \bar{\tau}_{j}\quad  \forall \,\tau_{i}, \tau_{j} \in \sigma(L). \label{normal-evals}
\end{align}
Since $z_{ii} = 0$ identically, there exist exactly $N$ zero eigenvalues. The mean density of eigenvalues is defined by the ensemble average of the density function
\begin{align}
\rho(z) = \left\langle \sum_{i = 1}^{N^{2}} \delta^{(2)}(z - z_{ij}) \right\rangle .
\end{align}
Apart from the $N$ exact zero modes, the eigenvalues lie inside a shifted circle defined parametrically by
\begin{align}
\frac{y^{2}}{4v^{2}} + \left( \frac{x}{2v} + 1\right)^{2} = 1, 	\label{normal-shape}
\end{align}
where $(x,y) \in \mathbbm{R}^{2}$ are coordinates on the real plane (see Fig.(\ref{fig:dissnormalevals})). 

\begin{figure}[htbp!]
\centering
\includegraphics[scale=0.7]{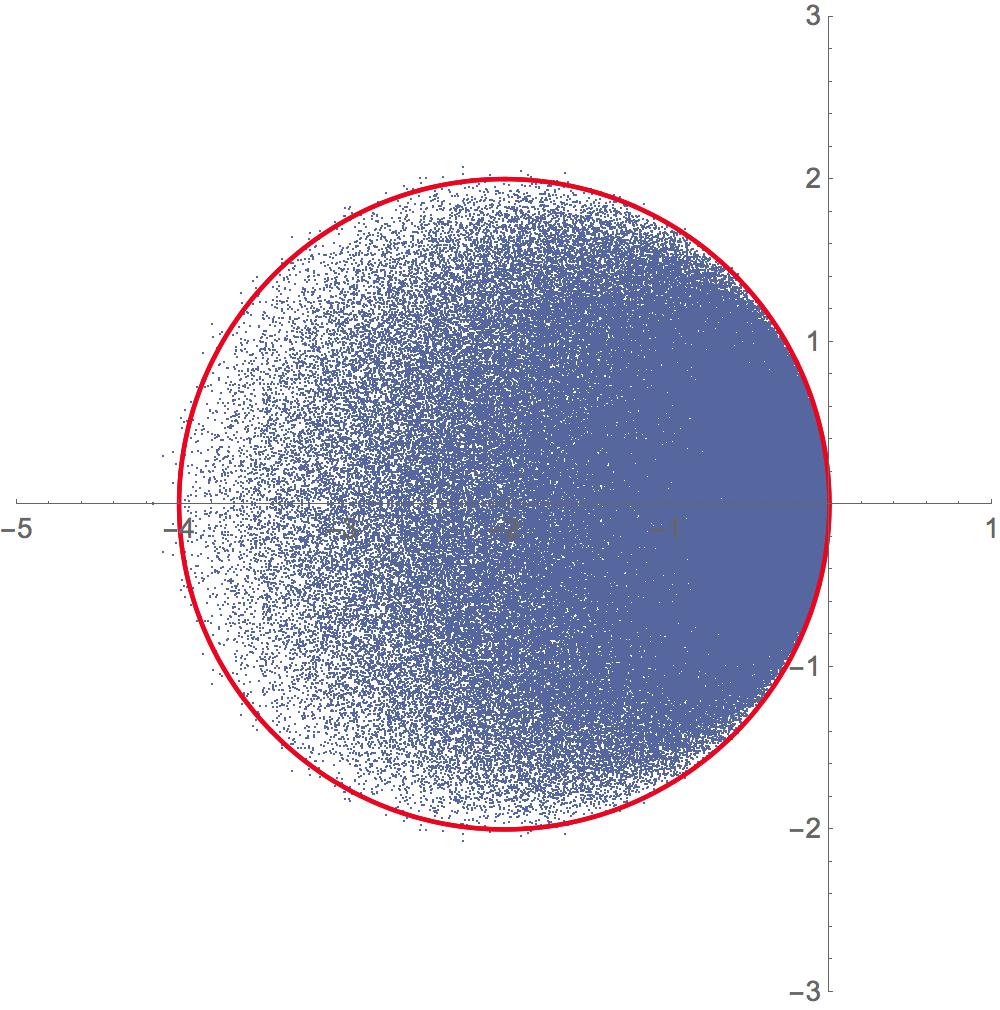}\\
\caption{\label{diss-normal-evals} Eigenvalues of simple dissipator with random normal jump operators, shown for a single realization with $v = 1$, $N  = 500$. The solid red line is the curve \eq{normal-shape}.  }
\la{fig:dissnormalevals}
\end{figure}

This demonstrates that the spectral gap is zero in the thermodynamic limit, which leads to the algebraic relaxation of Prop.(\ref{Result1b}). This result follows by direct computation, which is given in full detail in Appendix (\ref{normal}), and accomplished by utilizing the exact result for the $n$-point eigenvalue distribution functions
\begin{align}
R_{n}(\tau_{1}, .., \tau_{n}) = \det\left[ K(\tau_{i}, \tau_{j})\right]_{i,j = 1, ..., n}	,
\end{align}
where the kernel function is given by 
\begin{align}
 K(\tau_{i}, \tau_{j}) = \frac{N}{\pi v} \sum_{n = 0}^{N-1} \frac{(\tau_{i} \bar{\tau}_{j})^{n}}{n!} e^{ - \frac{N}{2v} ( |\tau_{i}|^{2} + |\tau_{j}|^{2})}.
\end{align}

In particular, the dissipative form factor is 
\begin{align}
F(t) &= \frac{1}{N^{2}} \left\langle \sum_{i,j} e^{ - t\left( |\tau_{i}|^{2}  + |\tau_{j}|^{2} - 2 \tau_{i}\bar{\tau}_{j}\right)} \right\rangle,\\
& = \frac{1}{N^{2}} \int d^{2} z_{1} d^{2} z_{2} \, \langle \rho(z_{1}) \rho(z_{2}) \rangle e^{ - t|z_{1}|^{2} - t|z_{2}|^{2} + 2 t \bar{z}_{1} z_{2}}, 
\end{align}
and the two-point density correlation function is
\begin{align}
 \langle \rho(z_{1}) \rho(z_{2}) \rangle = \langle \rho(z_{1})\rangle \delta^{(2)}(z_{1} - z_{2}) + R_{2}(z_{1}, z_{2}).
\end{align}
In the large $N$ limit, and setting $v = 1$, the form factor behaves as
\begin{align}
F(t) = \frac{(1 - e^{- t})^{2}}{t^{2}} + O(N^{-1}).\label{exact-normal-ff}
\end{align}
The asymptotic expansion up to order $O(1)$ is presented in Appendix (\ref{normal}). The power-law relaxation as $t^{-2}$ is consistent with the closing of the spectral gap.

\begin{figure}[htbp!]
\centering
\includegraphics[scale=0.7]{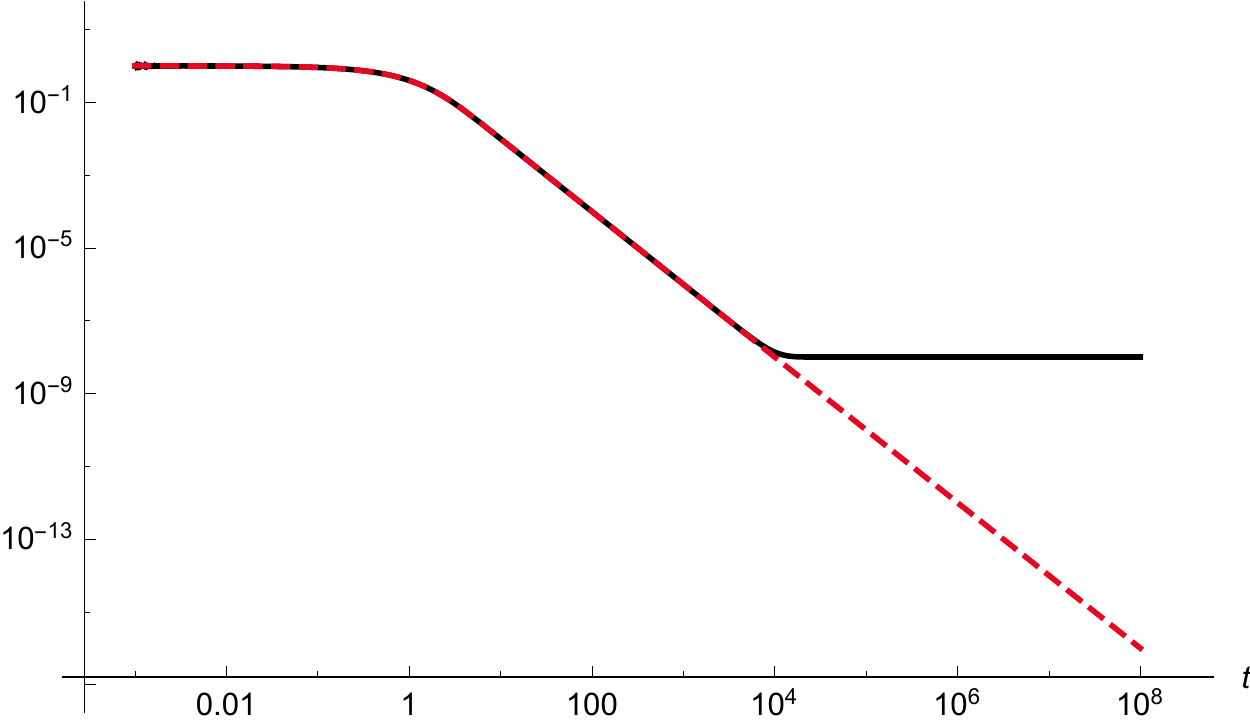}\\
\caption{\label{ff-normal-evals} Exact form factor \eq{ff-normal-exact} (black) and the leading order asymptotic formula \eq{exact-normal-ff} (dashed red) on a log scale for $N = 10^{8}$.}
\la{fig:ffnormal}
\end{figure}

In Fig.(\ref{fig:ffnormal}), we compare the exact form factor with the leading order asymptotic formula, to illustrate the regime in which the leading order behavior is relevant. The form factor appears to saturate around $\tau\sim \sqrt{N}$, before which it is well described by the large $N$ behavior. Since the asymptotic limit is known to be $F(\infty) = 1/N$, we expect that the asymptotic value takes over when $t^{-2} << N^{-1}$, i.e. $t> \sqrt{N}$. Note that, unlike the simple dissipator with Ginibre jump operators, the asymptotic value enters at the same order as the sub-leading correction to $F(t)$. Consequently, the cross-over regime observed in Fig.(\ref{fig:ffginibre}) is not observed in Fig.(\ref{fig:ffnormal}).

\subsection{Gaussian Unitary Ensemble}\label{sec:diss-gue}
\begin{figure}[htbp!]
\centering
\includegraphics[scale=0.7]{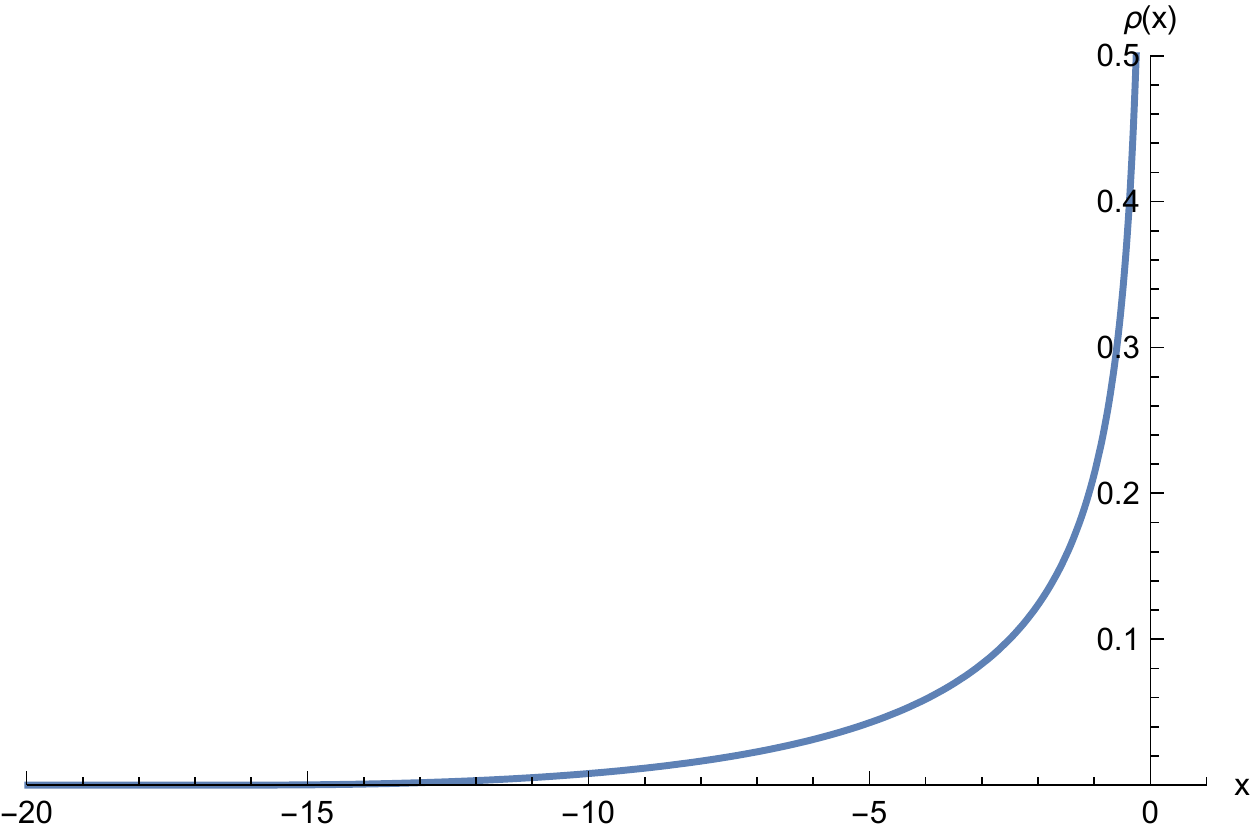}\\
\caption{ Density of dissipator eigenvalues which follows from the large $N$ resolvent (\ref{resolvent-dissipator}) with $v = 1$.}
\label{fig:dissipator-density}
\end{figure}

For Hermitian jump operators, the simple dissipator can be written as the square of the Liouville-von Neumann operator
\begin{align}
\mathcal{D} = - \left( L \otimes_{t} \mathbbm{1} - \mathbbm{1} \otimes_{t} L \right)^{2}, \label{herm-dissipator}
\end{align}
which acts on the density matrix as a nested commutator $\mathcal{D}\rho = - [ L, [L, \rho]]$. We work exclusively with the Gaussian unitary ensemble, in which $L_{ij} \in \mathbbm{C}$ with mean zero and variance $\langle \bar{L}_{ij}L_{kl}\rangle = \frac{v}{N} \delta_{ik}\delta_{jl}$. As in \eq{normal-evals}, the eigenvalues of the dissipator $z_{ij}$ can be written in terms of the eigenvalues of the jump operators, and are given by  
\begin{align}
z_{ij} = - (\tau_{i} - \tau_{j})^{2}	,
\end{align}
now being real-valued since $\tau_{i}\in \mathbbm{R}$ are the eigenvalues of an hermitian $L$. As with \eq{normal-evals}, there exist $N$ exact zero eigenvalues of the dissipator, along with a distribution on the real line which can be extracted from the resolvent
\begin{align}
G(z) = \frac{1}{N^{2}} \left\langle \sum_{i,j} \frac{1}{z + (\tau_{i} - \tau_{j})^{2}} \right\rangle ,	
\end{align}
which in the large $N$ limit evaluates to
\begin{align}
G(z) = -\frac{1}{2v} \left(1 - {}_{2}F_{1} \left( - \frac{1}{2}, \frac{1}{2} ; 2 ; -\frac{16 v}{z}\right) \right),\label{resolvent-dissipator}
\end{align}
with $N^{-1}$ corrections (see Appendix (\ref{sec:spec-LvN})). Obviously, this limit completely misses the $N$ zero eigenvalues, which will appear as a simple pole at $z = 0$ with residue $N^{-1}$. The density of eigenvalues is obtained from the resolvent $\rho(x) = -(1/\pi) {\rm Im}\, G(x + i 0^{+})$, and plotted in Fig. (\ref{fig:dissipator-density}). 

These results follow directly from the method of moments \cite{Wigner1955}. Writing the resolvent as a Laurent series for large $z$ away from the real axis gives
\begin{align}
	G(z) = \frac{1}{z} \sum_{k = 0}^{\infty} \frac{\mu_{k}}{z^{k}},\label{laurent-resolvent}
\end{align}
where we define the $k$-th trace moment of the GUE dissipator
\begin{align}
\mu_{k} = \frac{1}{N^{2}} \langle \tr \mathcal{D}^{k} \rangle = \frac{1}{N^{2}}\int \int dx\, dy \, \langle \rho(x) \rho(y)\rangle  \,  (x - y)^{2k}.
\end{align}
In the large $N$ limit, the two-point density correlation function can be approximated by the product of one-point densities. This approximation neglects the eigenvalue correlations. Defining $M_{k} =(-v)^{-k}\lim_{N\to \infty}  \mu_{k}$, we have the explicit expression

\begin{align}
M_{k} = \frac{1}{v^{k}} \int \int dx\, dy \, \rho_{sc}(x) \rho_{sc}(y) \,  (x - y)^{2k} \, = \frac{4^{2k+1} \Gamma(k+1/2) \Gamma(k+3/2)}{\pi \Gamma(k+2) \Gamma(k+3)},\label{moments-sc}
\end{align}
where the asymptotic mean density is the well-known semi-circle distribution
\begin{align}
	\rho_{sc}(x) = \frac{ 1}{2\pi v} \sqrt{4 v- x^{2}}, \quad - 2 \sqrt{v} \le x \le 2 \sqrt{v}, \quad \int \rho_{sc}(x) dx = 1.
\end{align}
The moments \eq{moments-sc} clearly obey the recursion relation 

\begin{align}
M_{k} = 16   \frac{(k-1/2) (k+1/2)}{(k+1)(k+2)} M_{k-1}	.\label{recursion-moments}
\end{align}
Inserting this into the Laurent series \eq{laurent-resolvent} and taking derivatives in order to massage the sum back to its original form, we arrive at the hypergeometric equation for the rescaled resolvent $F(\zeta) = - v G(1/\zeta)$, 
\begin{align}
(1 - 16  \zeta )\zeta F'' + (2 - 16 \zeta ) F'  + 4  F - 2 = 0,
\end{align}
which is solved by \eq{resolvent-dissipator}. 

The method of moments is also the most effective way to compute the dissipative form factor, which in this case can be compactly represented as as a hypergeometric function 

\begin{align}
F(t) = \,{}_{2}F_{2}\left(\frac{1}{2}, \frac{3}{2}; 2, 3; - 16 v t\right)	 + O(N^{-1}).
\end{align}
which relaxes algebraically at long times 

\begin{align}
F(t) \sim \frac{16}{3 \pi^{3/2} \sqrt{4 v \gamma }} \frac{1}{\sqrt{t}}.
\end{align}
The power-law in this case is expected since the spectral gap closes in the thermodynamic limit. However, the rate of relaxation is slower than both the complex Ginibre and random normal dissipators above. The technical reason for this is the $1/\sqrt{x}$ (integrable) divergence of the spectral density close to the origin. To see this, we may use the Pfaff transformation (\ref{pfaff}) to find the resolvent \eq{resolvent-dissipator} near the origin

\begin{align}
G(z) \to  -\left(\frac{1}{2 v} - \frac{4}{3\pi}\frac{\sqrt{z + 16 v}}{2 v \sqrt{z}}	\right),
\end{align}
which closely resembles the resolvent for the Mar\v{c}enko-Pastur distribution. The $1/\sqrt{x}$ divergence of the density thus becomes manifest, see Fig.(\ref{fig:dissipator-density}).

If we repeat the arguments in the previous section to estimate the relaxation time in terms of the Hilbert space dimension, we expect the large $N$ form factor to be valid for times $t^{-1/2} \ge N^{-1}$, which gives $\tau \sim N^{2}$ as approximately the relaxation time after which the dissipative form factor can be expected to saturate, identical to the case with complex Ginibre jump operators. The same algebraic decay was found in Ref. \cite{Tameshtit1993} for the decay of purity under Lindblad evolution in which the jump operator and the Hamiltonian are identical and drawn from the Gaussian orthogonal ensemble.

\section{Dissipators with Multiple Jump Operators}\label{sec:multi-diss}
In this section, we are concerned with the proof of Proposition (\ref{Result2}) for the long-time behavior of the propagator for the dissipator with multiple jump operators 
\begin{align}
\mathcal{D} = \sum_{a = 1}^{m} \mathcal{D}_{L_{a}}, \label{multi-diss-eq}
\end{align}
where $L_{a}$ are independent complex Ginibre matrices whose elements have zero mean and variance $\langle |L_{a,ij}|^{2}\rangle = v_{a}/N$. We assume $m$ is finite in the large $N$ limit. In Sec.(\ref{sec:gin-diss}), we proved Lemma (\ref{lemma}) for a simple dissipator. Here, we extend the argument to multiple jump operators, completing the proof. Let us introduce the Hermitian matrix 
\begin{align}
M = \sum_{a=1}^{m} L_{a}^{\dagger} L_{a},\label{M-def}
\end{align}
In terms of which the dissipator can be written
\begin{align}
\mathcal{D} = \sum_{a=1}^{m} 2 L_{a}\otimes_{t} L_{a}^{\dagger} - M\otimes_{t} \mathbbm{1} - \mathbbm{1} \otimes_{t} M\label{diss-multi-2}.
\end{align}
The leading order term in the $k^{th}$ moment of $\mathcal{D}^{k}$ will be order $N^{k+2}$. This obviously requires the diagrams to be planar, since a contraction with intersecting lines will reduce the number of closed loops. However, according to our graphical calculus, this requirement is not enough. A planar diagram which has contractions that connect the two edges must necessarily be subleading. We can see this inductively. For a single crossing, the edges which accomplish the crossing cannot participate in a closed loop, making it impossible to produce $k$ loops. Thus, such diagrams are at most order $N^{k+1}$. If there are two crossing contractions, then a closed loop can form which connects the edges; however, this closed loop then must involve two contractions, making it also order $N^{k+1}$ {\it at most}. With three crossing contractions, we have $N^{k}$, since out of the three crossing contractions we can construct at most one closed loop. Proceeding, we see clearly that for $p$ crossing contractions on the $k$-th moment, the diagrams are {\it at most} order $N^{k+2-p+ \lfloor p/2\rfloor }$, where $\lfloor p/2\rfloor$ is the largest integer $q \in \mathbbm{Z}$ such that $q \le p/2$ (in other words, if $p$ is even, $q = p/2$, otherwise $q = (p - 1)/2$). We conclude that the leading order diagrams are both planar {\it and} non-crossing. The insertion of any number of recycling terms will require a crossing contraction, and therefore be subleading to $N^{k+2}$. These arguments concern the topology of the diagrams which are possible for Ginibre jump operators, and are insensitive to the number of jump operators, and therefore apply to \eq{multi-diss-eq} for all $m$ (see Appendix (\ref{sec:top_expansion}) for an elaboration of this topological argument). The diagrammatics thus tells us that we will successfully capture the leading order trace moment by neglecting the recycling term and taking the $N \to \infty$ limit of the non-crossing truncation of the dissipator
\begin{align}
\frac{1}{N^{2}} \left\langle \tr \mathcal{D}^{k}\right\rangle =  \lim_{N \to \infty} \frac{(-1)^{k}}{N^{2}} \left\langle \left( M \otimes_{t} \mathbbm{1} + \mathbbm{1}\otimes_{t} M \right)^{k} \right\rangle + O(N^{-1}).\label{moments-multi-diss}
\end{align}

This completes the proof of Lemma (\ref{lemma}) for the case of multiple jump operators. The limit which appears on the right-hand side is necessary, since the moments of the dissipator are not strictly equivalent to the moments of the non-crossing truncation of the dissipator at subleading order. As we emphasized before, this is essentially a trick which allows us to count the  planar diagrams for which the two edges are disconnected.

As the sum of $m$ independent random Wishart matrices, we may find the eigenvalue density of $M$ using the methods of free probability theory. This requires first computing the self-energy $\Sigma$ (known as the R-transform in free probability theory, see Theorem 18 of Ref \cite{Mingo2017}) of the single Wishart matrix $L^{\dagger}L$ (which we do not normalize by $N$). By inverting the resolvent Eq.\eq{resolvent-MP}, we find for the single Wishart matrix

\begin{align}
z[G] & = \frac{1}{ G (1 - v G)},
\end{align}
from which we find the self-energy by definition 
\begin{align}
\Sigma[G] = z[G] - \frac{1}{G} = 	\frac{v}{1 - v G} .
\end{align}
The self-energy for a sum of $m$ Wishart matrices with different $v_{a}$ is given by the sum of self-energies (which is the famous result that the $R$-transform is additive for free random variables)

\begin{align}
\Sigma_{m}[G] = \sum_{a=1}^{m} \frac{v_{a}}{1 - v_{a} G}.
\end{align}
The resolvent for $M$ then follows implicitly by the equation

\begin{align}
G_{m}(z) = \frac{1}{z - \Sigma_{m}[G_{m}(z)]}.	
\end{align}
We consider two scenarios for simplicity below. 

\bigskip
\paragraph*{Two Distinct Jump operators}

Here we consider the case with two jump operators with variances $v_{1}  = 1$ and $v_{2} =\gamma$. The self-energy is
\begin{align}
\Sigma_{2}[G] = \frac{1}{1 - G} + \frac{\gamma}{1 - \gamma G}.
\end{align}

The resulting resolvent has a spectral gap which scales linearly in $\gamma$ for small $\gamma$ and tends to a constant as $\gamma \to \infty$. This follows from an obvious duality in the eigenvalue density for $M = L_{1}^{\dagger}L_{1} +  L_{2}^{\dagger}L_{2}$
\begin{align}
\rho_{\gamma}(x) = \frac{1}{\gamma} \rho_{1/\gamma}(x/\gamma)	.\label{density-duality}
\end{align}

\begin{figure}
\centering
\includegraphics[scale=1.1]{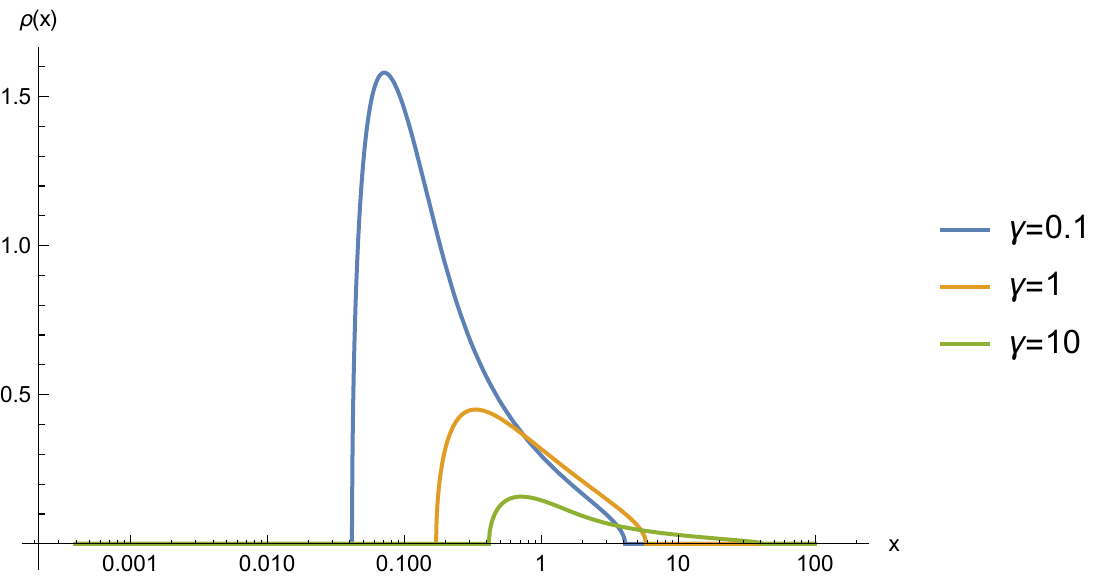}\\
\caption{Eigenvalue density $\rho(x)$ for $M$ \eq{M-def} for $m = 2$ with variances $v_{1} = 1$ and $v_{2} = \gamma$, presented on a semi-log plot in order to clearly see the spectral gap. Under the rescaling \eq{density-duality}, the $\gamma = 0.1$ curve will collapse to the $\gamma = 10$ curve.  $\gamma = 1$ represents a self-dual point, and is described by Eq.(\ref{multi-diss-density-m}) for $m = 2$.}
\label{fig:multi-diss-1}
\end{figure}

The small $\gamma$ gap follows easily from first order perturbation theory, and scales linearly in $\gamma$, which implies that $\rho_{\gamma}(x) = 0$ for $x < \delta \sim \gamma$. Therefore, at large $\gamma$, the dual density $\rho_{1/\gamma}(x/\gamma) = 0$ for $x/\gamma < \delta \sim 1/\gamma$. Therefore, $\rho_{\gamma}(x)$ vanishes for $x < \Delta \sim 1$ for large $\gamma$. A plot of the leading order eigenvalue density for various $\gamma$ is presented in Fig.(\ref{fig:multi-diss-1}). 

The existence of a spectral gap in $M$ implies exponential decay of the dissipative form factor. We explore this in more detail in the next section. 

\bigskip
\paragraph*{Identical Jump Operators} Now we consider $m$ jump operators with the same variance $v_{a} = v$. The self-energy is
\begin{align}
\Sigma_{m} = \frac{m v}{1 - v G}	,
\end{align}
which allows us to find a simple expression for the holomorphic resolvent
\begin{align}
G_{m}(z) =\frac{ z - v (m-1) 	- \sqrt{ (z - v (m-1 ))^{2} - 4 v z}}{2 v z}.\label{multi-diss-resolvent-m}
\end{align}
The density of eigenvalues in terms of the dimensionless variable $\xi = x/v$ is
\begin{align}
\rho(\xi) = & = \frac{1}{2\pi v \xi} \sqrt{(\xi_{+} - \xi)(\xi - \xi_{-})}, \quad \xi_{-} \le \xi \le 	\xi_{+}, \quad \xi_{\pm} = (1 \pm \sqrt{m})^{2}.\label{multi-diss-density-m}
\end{align}
The leading order contribution to the moment generating function for the dissipator is 
\begin{align}
F(t) & = \left(\int_{\xi_{-}}^{\xi_{+}}  d \xi  \frac{1}{2\pi  \xi} \sqrt{(\xi_{+} - \xi)(\xi - \xi_{-})} e^{ - v t \xi} \right)^{2},\\
& = e^{ - 2v  \xi_{-}t }\left(\int_{0}^{4 \sqrt{m}} \frac{d u}{2\pi(u+\xi_{-})} \sqrt{ (4 \sqrt{m} - u) u} e^{ - v t u}\right)^{2}. \label{multi-diss-ff1}
\end{align}
At large times, the integral is dominated by values of $u t \sim O(1)$. Defining the new variable $u' = ut$, we get the asymptotic expression for large $t$
\begin{align}
F(t) & \sim  e^{ - 2v  \xi_{-}t } \left(\int_{0}^{\infty} \frac{1}{t^{3/2}}\frac{d u'}{2\pi(\xi_{-})} \sqrt{ (4 \sqrt{m}) u'} e^{ - v u'}\right)^{2},\nonumber\\
& = e^{ - 2 v \xi_{-} t} \frac{1}{t^{3}} \frac{4 \sqrt{m}}{(2\pi \xi_{-})^{2}} \left( \int du'\, \sqrt{u'} e^{ - v u'} \right)^{2}=  C t^{-3}  e^{ - 2 v \xi_{-} t}, \quad C = \frac{ \sqrt{m}}{4\pi  v^{3} (\xi_{-})^{2} },\label{multi-diss-ff}
\end{align}

which completes the proof of Prop.(\ref{Result2}), and gives the explicit formula for the constant coefficient $C$.

\begin{figure}
\centering
\includegraphics[scale=0.7]{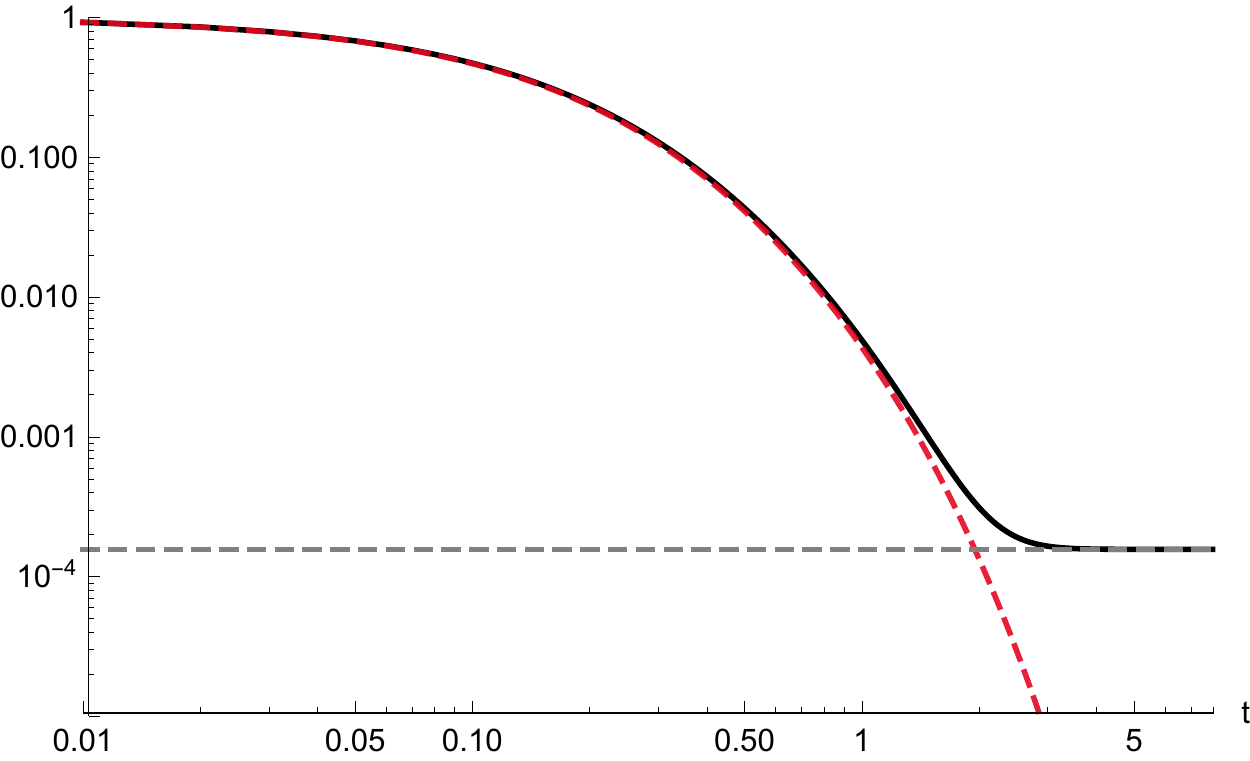}\\
\caption{Numerically exact form factor (black), the leading order asymptotic expression (\ref{multi-diss-ff1}) (dashed red), and the asymptotic value $F(\infty) = 1/N^{2}$ (dashed gray), shown for a single realization with $N = 80$, $m = 4$ and $v = 1$. }
\label{fig:gap-nh}
\end{figure}

\section{Simple Lindbladian}\label{sec:ham-diss}

In this section, we provide the proof of Proposition (\ref{Result3}) and elaborate on the structure of the spectral gap. We consider here the simple Lindblad superoperator
\begin{align}
	\mathcal{L} = - i \left( H \otimes_{t} \mathbbm{1} - \mathbbm{1} \otimes_{t} H\right) + \gamma \left( 2 L \otimes_{t} L^{\dagger} - L^{\dagger} L \otimes_{t} \mathbbm{1} - \mathbbm{1} \otimes_{t} L^{\dagger}L \right),
\end{align}
where $H$ is a complex hermitian random matrix drawn from the Gaussian unitary ensemble with zero mean and variance $\langle \bar{H}_{ij} H_{kl} \rangle = N^{-1} \delta_{ik}\delta_{jl}$, the jump operator $L$ is a complex Ginibre random matrix with variance $\langle \bar{L}_{ij} L_{kl}\rangle = N^{-1} \delta_{ik}\delta_{jl}$, and the dissipative coupling $\gamma > 0$. In terms of the non-Hermitian Hamiltonian $K = H - i \gamma L^{\dagger}L$, the Lindblad superoperator takes the form 
\begin{align}
	\mathcal{L} =  \tilde{\mathcal{L}}  + 2\gamma   L \otimes_{t} L^{\dagger} , \quad \tilde{\mathcal{L}} = -i \left( K\otimes_{t} \mathbbm{1} -   \mathbbm{1} \otimes_{t} K^{\dagger}\right). \label{eff-ham}
\end{align}
We will refer to $\tilde{\mathcal{L}}$ as the non-unitary LvN generator (which is the non-crossing truncation per Def.(\ref{noncrossing})), while the second term is the recycling term. Lemma (\ref{lemma}) implies that the moments of the Lindbladian are to leading order given by the diagrams with disconnected legs, which follows in a way identical to (\ref{moments-multi-diss})
\begin{align}
\frac{1}{N^{2}}\left\langle \tr \mathcal{L}^{n} \right\rangle 	 = \lim_{N \to \infty} \frac{1}{N^{2}} \left\langle \tilde{\mathcal{L}}^{n} \right\rangle + O(N^{-1}).
\end{align}
One can construct a biorthogonal basis for $K$ using right-eigenvectors satisfying $K | R_{i}\rangle = z_{i} | R_{i}\rangle$, and left eigenvectors $\langle L_{i}| K = z_{i} \langle L_{i}|$, such that $\langle L_{i}|R_{j}\rangle = \delta_{ij}$. The conjugate transpose of the right eigenvectors satisfy $ ( K |R_{i}\rangle )^{\dagger} = \langle R_{i}| K^{\dagger} = \bar{z}_{i} \langle R_{i}|$. Using these facts, it is easy to see that the eigenmodes of the non-unitary LvN superoperator are given by $| R_{i}\rangle \langle R_{j}|$ with the spectrum
\begin{align}
\sigma(\tilde{\mathcal{L}} ):=\Big\{ - i \left(z_{i} - \bar{z}_{j}\right): \quad z_{i} \in \sigma(K) \Big\}.
\end{align}
From this, we find that the moment generating function for the simple Lindbladian is given to leading order by
\begin{align}
F(t) &= \lim_{N \to \infty} \frac{1}{N^{2}}\left\langle  \tr \, e^{t \tilde{\mathcal{L}} } \right\rangle 	+ O(N^{-1}),\\
& = \int d^{2} z d^{2} \zeta   \, e^{ - i (z - \bar{\zeta}) t} \rho_{K}(z) \rho_{K}(\zeta)  = \left| \int d^{2} z e^{ - i z t} \rho_{K}(z) \right|^{2},\label{ff-nh}
\end{align}
where the mean density for the non-hermitian Hamiltonian is defined as usual by
\begin{align}
\rho_{K}(z) = \lim_{N \to \infty} \frac{1}{N} \left\langle \sum_{i = 1}^{N} \delta^{(2)}(z - z_{i}) \right\rangle .
\end{align}
The form factor \eq{ff-nh} was also considered in Ref. \cite{Gudowska-nowak1998}, where it was interpreted as a survival probability for non-unitary LvN evolution. In \cite{Alhassid1998}, it was calculated for a related effective Hamiltonian with a lower-rank non-hermitian deformation in the weakly open limit, and related to the spectral autocorrelation function describing the induced photodissociation of molecules. Using the results in Appendix (\ref{sec:nonhermitian-ham}), the moment generating function for the density can be written
\begin{align}
M(t) = \int d^{2} z \, e^{ - i z t} \tilde{\rho}(z) = \int dx \int_{-y(x)}^{y(x)} dy  \, \rho(x) e^{ - (x + i y) t},
\end{align}
where the boundary of the support of the density is described by the curve
\begin{align}
y^{2} = \frac{4}{\gamma x} - 	\left[ \frac{1}{x} - \frac{\gamma}{ (1 + \gamma x)}  + \frac{1}{\gamma} \right]^{2}, \label{ycurve}
\end{align}
and the density contained in this curve is independent of $y$ and given by
\begin{align}
\rho(x) = \frac{1}{4\pi} \left[ \frac{1}{x^{2}} - \frac{\gamma^{2}}{(1 + \gamma x)^{2}} + 1 \right].\label{resonance-density}	
\end{align}
Note that we have switched to the convention used in Appendix (\ref{sec:nonhermitian-ham}), where we computed the density for the ``rotated" matrix $i K$. Since the spectrum of $i K$ is symmetric about the $x$ axis, the spectral gap $\delta$ is given by the smallest real-valued $x_{min}$ which solves \eq{ycurve} when $y= 0$. The resulting cubic equation for $x$ has explicit solutions, though their complicated form is not illuminating. In Fig.(\ref{fig:gap-nh}), $x_{min}(\gamma)$ is plotted revealing a non-monotonic function of the dissipative coupling. The spectral gap is known to close in both the $\gamma = 0$ and $\gamma \to \infty$ limits, and obtains a maximal value when $\gamma = O(1)$. The asymptotic behavior of the gap can be easily determined in the small $\gamma$ limit from the solution to
\begin{align}
\frac{2}{\gamma x } \sim \left( \frac{x+ \gamma}{x \gamma}\right)^{2} \Rightarrow x_{min}\sim \gamma ,	
\end{align}
whereas for large $\gamma$, we can insert the scaling ansatz $x = \gamma^{- \alpha}$, and find
\begin{align}
x_{min} \sim (4 \gamma)^{-1/3}	.
\end{align}
\begin{figure}
\centering
\includegraphics[scale=0.7]{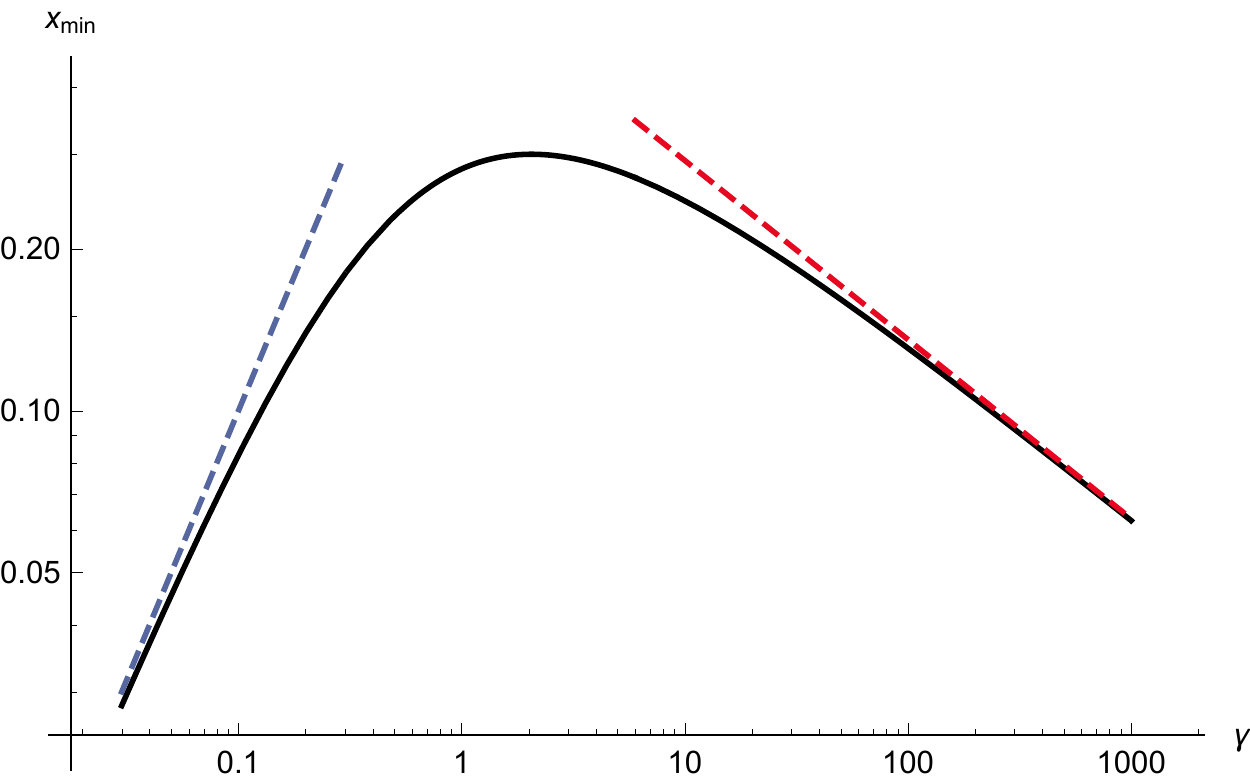}\\
\caption{Spectral gap $x_{min}(\gamma)$ of non-Hermitian Hamiltonian as a function of dissipative coupling $\gamma$ shown on a log scale (solid black), along with the asymptotic linear growth $x_{min}=\gamma$ (dashed blue) and power-law decay $x_{min} =(4\gamma)^{-1/3}$ (dashed red).}
\label{fig:gap-nh}
\end{figure}
To obtain the large $t$ asymptotics, we first perform the $y$ integration for the moment generating function
\begin{align}
M(t) = \frac{2}{t} \int_{x_{min}}^{x_{max}} dx\, \rho(x)\, e^{ - x t} \sin(y(x) t), \label{mom-lind}
\end{align}
where the lower limit is set by the spectral gap, while the upper limit is the largest real-valued solution to \eq{ycurve} with $y = 0$. This has the obvious upper bound

\begin{align}
M(t) \le \frac{2}{t} \int_{x_{min}}^{x_{max}} dx\, \rho(x)\, e^{ - x t},
\end{align}

which leads directly to an upper bound for the form factor 

\begin{align}
F(t) \le \left(\frac{2}{t} \int_{x_{min}}^{x_{max}} dx \rho(x) e^{ - x t} \right)^{2}.\label{ff-bound2}
\end{align}

The integral can be evaluated directly using the exact form of the resonance width distribution \eq{resonance-density}
\begin{align}
\int_{x_{min}}^{x_{max}} dx \rho(x) e^{ - x t} = \frac{1}{4\pi} \left( I_{x_{max}}(t) - I_{x_{min}}(t)\right),
\end{align}
where the indefinite integral is given by 
\begin{align}
I_{b}(t) & = -e^{ - b t} \left[ \frac{1}{t}   + t\,e^{ b t} {\rm Ei} ( - b t)+ \frac{1}{b}  - t\, e^{ (\gamma^{-1} +  b )t} {\rm Ei} \left( - (\gamma^{-1} +  b) t\right)  - \frac{1}{\gamma^{-1} + b} \right].
\end{align}

At large time, we may use the definition of the exponential integral and its asymptotic behavior 
\begin{align}
{\rm Ei}(- x) =- \int_{x}^{\infty}	\frac{e^{ - y}}{y} dy, \quad {\rm Ei}(-x) \to - \frac{e^{ - x}}{x}\quad{\rm as}\,\, x \to \infty,
\end{align}
to find the leading behavior of $I_{b}(t)$ at long times
\begin{align}
I_{b}(t) \to - \frac{e^{ - b t}}{t}.
\end{align}
This shows that $I_{x_{min}}(t)$, which has a slower decay rate, will be dominant at late times. Then plugging this into the expression \eq{ff-bound2} leads to the result stated in Prop.(\ref{Result3}),
\begin{align}
F(t) \le \left( \frac{1}{2\pi t^{2}} e^{ - \delta t} \right)^{2} = \frac{1}{4\pi^{2} t^{4}} e^{ - 2\delta t},
\end{align}
where $\delta = x_{min}$ is the spectral gap of $K$. This result is strictly valid only for $\gamma \ne \{0, \infty\}$, since the limits $t \to \infty$ and $\gamma \to 0\, (\infty)$ do not commute in the expression for $I_{b}(t)$.

\begin{figure}
\centering
\includegraphics[scale=0.7]{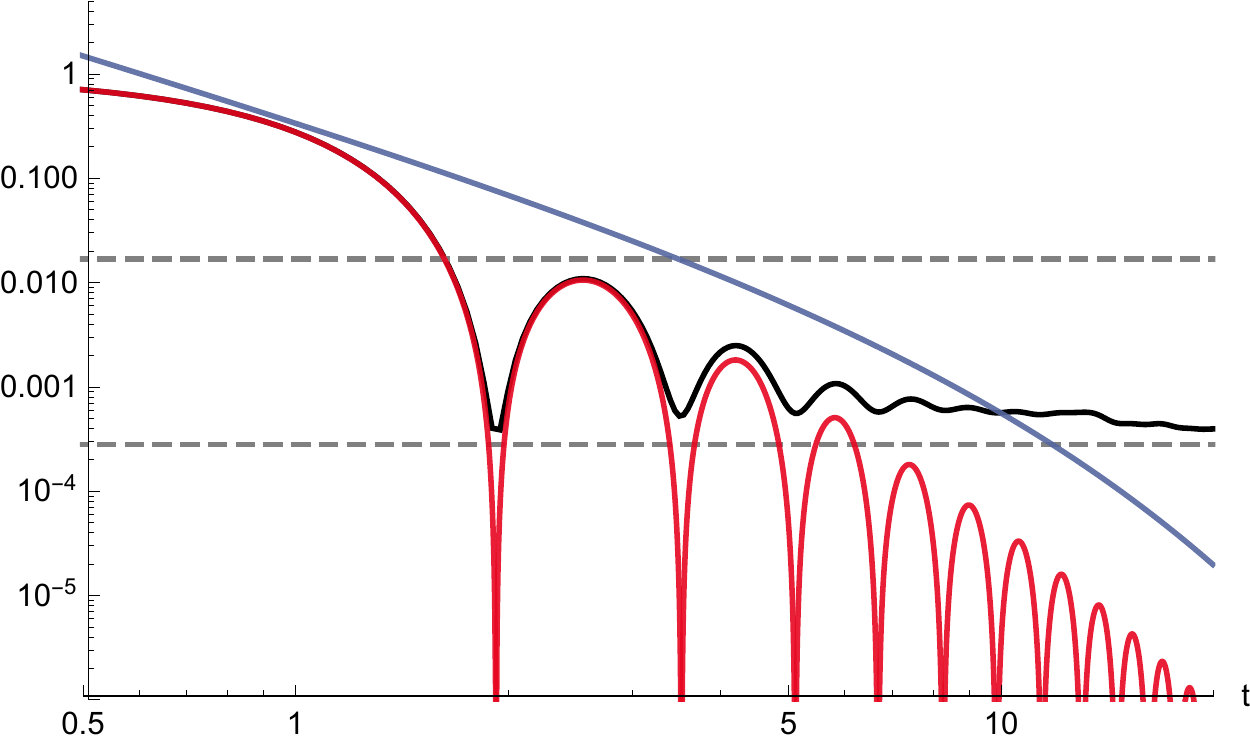}\\
\caption{ Numerically exact dissipative form factor (black), the leading order asymptotic formula which is the square of (\ref{mom-lind}) (red), and the exact upper bound (\ref{ff-bound2}) (blue), plotted on a log scale for $N = 60$, $\gamma = 0.1$, averaged over $100$ samples. The horizontal dashed lines are at $1/N$ (the order of the first subleading correction) and $1/N^{2}$ (the asymptotic limit $F(\infty)$). }
\la{fig:ffsimplelindblad}
\end{figure}

We compare our asymptotics and upper bound with the exact numerical form factor in Fig.(\ref{fig:ffsimplelindblad}). We observe that the exact form factor never seems to dip below its asymptotic value at any intermediate times, even though the large $N$ asymptotic formula (\ref{mom-lind}) does hit zero. Increasing $N$ in Fig.(\ref{fig:ffsimplelindblad-2}) shows a clear trend of the first minimum in the form factor getting deeper with increasing $N$. This gives support to the conclusion that in the asymptotic limit, the form factor does vanish at finite time of order $\Delta^{-1}$ and determined solely by the dissipative coupling. The figure also indicates the regimes where the large $N$ asymptotics fails in finite-sized systems. This is approximately when the first subleading $O(N^{-1})$ correction takes over, and occurs before the leading asymptotic formula dips fully below $F(\infty)$. The asymptotic formula is an excellent fit to the exact numerical $F(t)$ for $F(t) > 1/N$, and compares favorably for intermediate times when $N^{-1} > F(t) > N^{-2} = F(\infty)$.

\begin{figure}
\centering
\includegraphics[scale=1.]{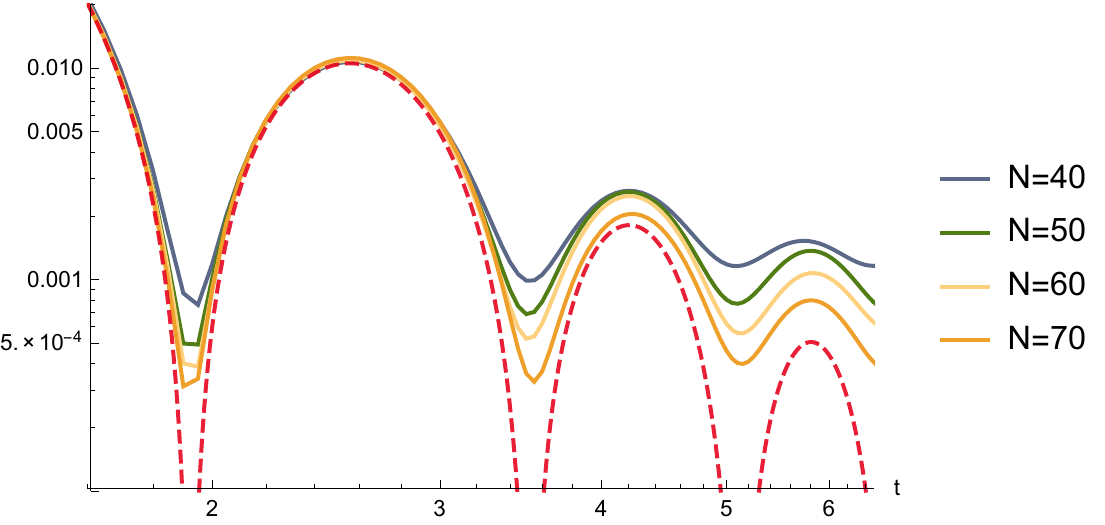}\\
\caption{ Numerically exact form factor $F(t)$ for various $N$ approach the leading order asymptote (dashed red) with increasing $N$, shown on a log plot.}
\la{fig:ffsimplelindblad-2}
\end{figure}

\section{Conclusion}

In this paper we have introduced the dissipative form factor and determined its asymptotic long-time behavior for various ensembles of random Lindblad generators. The picture that emerges is that exponential relaxation to steady state is generic and to be expected when the dynamics is governed by multiple non-commuting operators. We see this in particular for a dissipator with multiple jump operators and a simple Lindbladian with a Hamiltonian and a single jump operator. We also find that random simple dissipators are gapless, and lead to algebraic relaxation of observables and correlation functions in the thermodynamic limit.

Our work is in the spirit of Wigner's original proposal for a statistical description of complex atomic nuclei. Our setting extends this work to the time domain and asks how complex open systems equilibrate. An interesting follow-up to our work would be to characterize the structure of the steady state to which the random Lindblad equation flows, and how this depends on the spectral gap. A thermodynamic closing of the spectral gap has been implicated in second-order dissipative phase transitions \cite{Minganti2018}. In this context, it would be interesting to study the simple Ginibre dissipator which has a non-trivial steady state and is also gapless in the thermodynamic limit.

The closing of the spectral gap for the simple dissipator requires deeper understanding. Even though the evolution is dissipative for such a generator, the closing of the spectral gap indicates the existence, in the thermodynamic limit, of very long lived modes. This suggests that some sort of integrability emerges. The relaxation times involved in such systems (except for the random normal jump operators) implicate simple diffusion, though the lack of spatial structure perhaps produces more questions than answers along this line of reasoning. 

As it stands, we believe the random Lindbladian can serve a benchmark for studies of integrability and thermalization in open quantum systems. In a closed chaotic quantum system, the spectral correlations are reproduced by random matrix theory, indicating universality in quantum chaotic systems. To understand how these features might come to bear on questions of universality of Lindblad evolution, we should understand how to interpret the jump operators. One possibility which we find promising is to invoke the eigenstate thermalization hypothesis (ETH) for thermalizing systems. In particular, it was originally argued that observables in closed thermalizing systems can be treated as random matrices \cite{Srednicki1999}. Since the jump operators can be constructed from the observables on the Hilbert space, the ETH would posit the jump operators as being described by random matrices in the energy eigenstate basis. This motivates our speculation that random Lindblad equations provide a universal dynamical description of open chaotic quantum systems. Presumably the signatures of universality  will appear in the subleading contributions to the dissipative form factor, as they do for the spectral form factor for closed systems. This presents a promising direction for future work.

\bigskip
{\it Note Added - } While this paper was in preparation, the preprint \cite{Denisov2018} appeared which considered the spectrum of pure dissipators with the maximal $N^{2} - 1$ number of jump operators. They similarly found a spectral gap, as well as an explicit expression for the limiting large $N$ distribution of eigenvalues.

\bigskip
{\it Acknowledgements}\quad  I thank my collaborators on \cite{can}, out of which this work grew: S. Gopalakrishnan, V. Oganesyan and D. Orgad, with whom I have shared many fruitful discussions. The main results here and in Ref.\cite{can} were announced on October 4, 2018 in the workshop ``Random Matrices, Integrability, and Complex Systems" in Yad Hashmona, Israel. I have also benefited from discussions with A. Abanov, V. Albert, J. Feinberg, T. Seligman, W. Tarnowski, and V. K. Varma.

\appendix

\section{Asymptotic Decay Rate and Spectral Gap}\label{sec:replica}

In this appendix, we provide an argument to support the connection between our results on the asymptotic decay rate and the existence of a spectral gap. For quenched ``disorder", the spectral gap can be obtained formally from the propagator $K_{t}$ from the procedure

\begin{align}
\Delta = - \lim_{t \to \infty}	\frac{1}{t} \log \hat{F}(t), \quad \hat{F}(t) = \frac{1}{N^{2}} \left( {\rm tr} \, K_{t} - 1\right).
\end{align}

This definition is sufficient for our purposes since the steady state is unique and purely imaginary eigenvalues occur with vanishing probability. Note that the form factor coincides with the ensemble average of $\hat{F}(t)$ in the large $N$ limit. The extreme value statistic would then require an ensemble averaging of this quantity

\begin{align}
\langle \Delta \rangle = - \left\langle\lim_{t \to \infty}	\frac{1}{t}  \log \, \hat{F}(t) \right\rangle .
\end{align}

We can only proceed by postulating that the infinite time limit commutes with ensemble averaging. This leaves us to evaluate $\langle \log \hat{F}\rangle$, which can be done formally by a replica-like limit

\begin{align}
\left\langle \log \, \hat{F}(t) \right\rangle  = \lim_{n \to 0} \frac{1}{n} \left( \langle \hat{F}^{n} \rangle - 1\right).
\end{align}

The expectation value will now involve moments of the trace of the propagator. To understand the structure of this, let us consider $\langle \hat{F}^{2} \rangle$. Expanding the exponential, we see that this object will involve moments of the form

\begin{align}
\left\langle \frac{1}{N^{2}} \tr \mathcal{L}^{p} \, 	\frac{1}{N^{2}} \tr \mathcal{L}^{q}\right\rangle .
\end{align}

Since a product of traces is involves, the basic starting point for constructing the diagrams for this will be four disconnected loops. In other words, replacing both instances of $\mathcal{L}$ with the identity produces four disconnected loops, each giving a factor of $N$. Insertions of the Hamiltonian or jump operators and subsequent Wick contraction will cut these loops and either join two loops, reducing the total order of the diagram, or split one loop into two, increasing the order. Following the arguments given in Appendix \ref{sec:top_expansion}, this implies that the leading order diagrams will be given by neglecting Wick contractions which bridge the two moments of $\mathcal{L}$. 

We can put this in slightly different language, more familiar in the statistical mechanics setting. Let $\langle \rho(z_{1}) \rho(z_{2}) \rangle$ be the two-point eigenvalue density correlation function. Then

\begin{align}
\left\langle \frac{1}{N^{2}} \tr \mathcal{L}^{p} \, 	\frac{1}{N^{2}} \tr \mathcal{L}^{q}\right\rangle  = \int \, z_{1}^{p}\, z_{2}^{q} \langle \rho(z_{1}) \rho(z_{2}) \rangle d^{2} z_{1} d^{2} z_{2}.
\end{align}

In the large $N$ limit, our diagrammatic argument essentially boils down to the statement that the connected two-point function is subleading in $N$, i.e.

\begin{align}
\langle \rho(z_{1}) \rho(z_{2}) \rangle = 	\langle \rho(z_{1})\rangle \langle  \rho(z_{2}) \rangle  + O(N^{-1}).
\end{align}

Therefore, returning to the object of interest, we have that to leading order

\begin{align}
\left\langle \log \, \hat{F} \right\rangle  =  \lim_{n \to 0} \frac{1}{n} \left( \langle \hat{F} \rangle ^{n} - 1\right) + O(N^{-1}) = \log \langle \hat{F} \rangle  + O(N^{-1}).
\end{align}

This suggests that the trace of the propagator is self-averaging. Note that now we have only to take the logarithm of the dissipative form factor. Therefore, we may extract the average spectral gap via

\begin{align}
\langle \Delta \rangle \approx - \lim_{t \to \infty} \, \frac{1}{t} \log F(t).
\end{align}

Though not a rigorous proof, this argument supports the identification made in the paper between the asymptotic decay rate and the spectral gap.




\section{Complete Diagrams for Second Moment of the Simple Dissipator}

\begin{figure}[htbp!]
\centering
\includegraphics[scale=0.6]{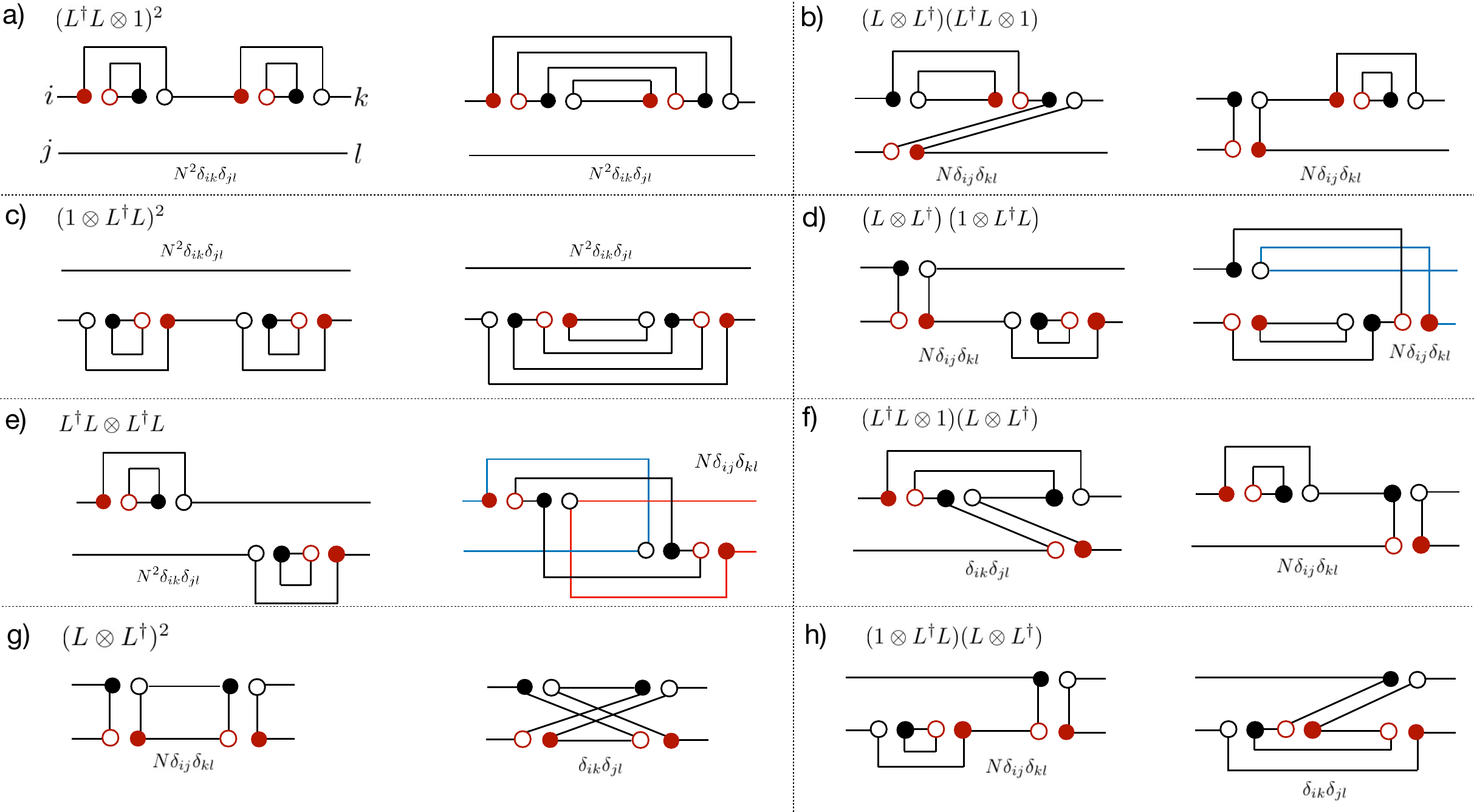}\\ 
\caption{\label{diagram-dissipator-complete} Diagrams for the second moment of the dissipator with a single complex Ginibre jump operator. All diagrams are multiplied by the square of the variance $(v/N)^{2}$.}
\la{fig:diagrams-second-order-complete}
\end{figure}

The diagrams in (\ref{fig:diagrams-second-order-complete}) can be used to compute the second moment (\ref{second-moment-diss}) exactly. These results can also be used to compute the third moment

\begin{align}
\left\langle \mathcal{D}^{3}_{ijkl}\right\rangle = (-v)^{3} \left( 22 + \frac{8}{N^{2}}\right) \left( \delta_{ik} \delta_{jl} - \frac{1}{N} \delta_{ij}\delta_{kl}\right).	
\end{align}

The leading order coefficient multiplying $(\delta_{ik}\delta_{jl} - N^{-1}\delta_{ij}\delta_{kl})$ can be expressed using moments of the Wishart matrix $L^{\dagger}L$ via

\begin{align}
(-1)^{n} \sum_{q= 0}^{n} \left( { n \atop q}\right) \langle \tr (L^{\dagger} L)^{n-q} \rangle \langle \tr (L^{\dagger}L)^{q}\rangle,
\end{align}
which can be computed using the Mar\v{c}enko-Pastur distribution \eq{density-MP}
\begin{align}
\langle \tr (L^{\dagger}L)^{n}\rangle 
& = \int_{0}^{4v} \, dx \, x^{n} \rho_{w}(x) = \frac{2(4v)^{n} }{\pi} \int_{0}^{1} dy \, y^{n-1/2} \sqrt{1-y}, \\
&= \frac{2 (4 v)^{n}}{\pi}  B(n+1/2, 3/2),
\end{align}

where $B(x,y)$ is the Euler beta function. 

\section{Topological Expansion and Proof of Lemma (\ref{lemma})}\label{sec:top_expansion}

\begin{figure}[htbp!]
\centering
\includegraphics[scale=0.6]{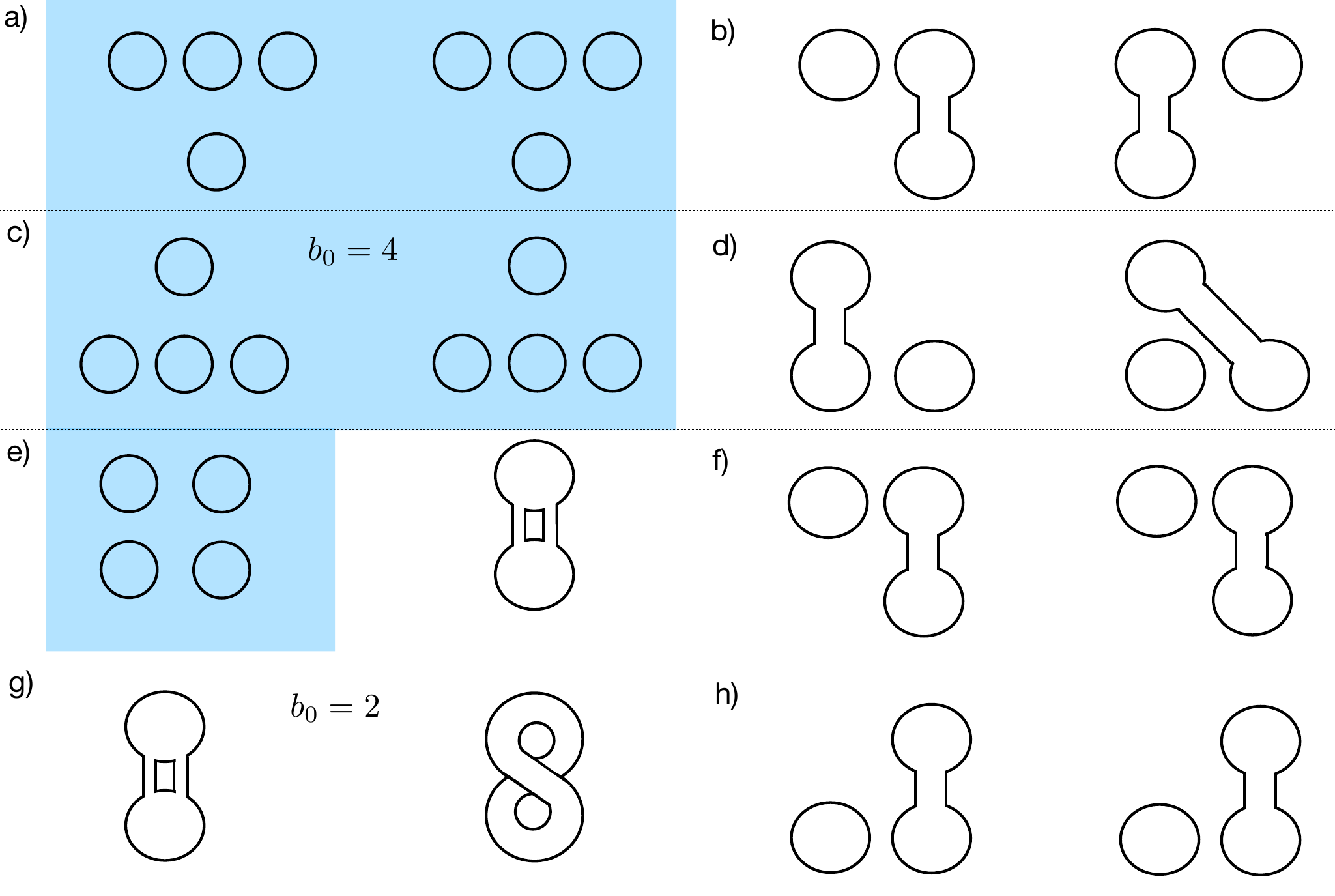}\\ 
\caption{\label{graphs-top-complete} Taking the trace of the open-legged diagrams in Fig.(\ref{fig:diagrams-second-order-complete}) produces graphs topologically equivalent to those presented here, labeled in the table as in Fig.(\ref{fig:diagrams-second-order-complete}). Leading order graphs with $b_{0} = 4$ are highlighted in blue, while the rest, all with $b_{0} = 2$, are subleading. The zeroth Betti number of each graph is indicated. Note: despite appearances, both graphs in (g) are topologically equivalent.}
\la{fig:graphs-top-complete}
\end{figure}

The moments of the simple dissipator enjoy a topological expansion

\begin{align}
M_{n} = \frac{1}{N^{2}} \left\langle {\rm tr} \, \mathcal{D}^{n} \right\rangle =  \frac{v^{n}}{N^{n+2}} \sum_{b_{0}} n_{b_{0}} N^{b_{0}}	
\end{align}

The factor $n_{b_{0}}$ is a combinatorial factor which counts the number of closed loops $b_{0}$, equal to the zeroth Betti number of the resulting graph. This coefficient must also account for the factors of $2$ or $-1$ coming from the different terms in the dissipator. Let us consider integer partitions of $n = p + q + r$. Then we may write this as

\begin{align}
M_{n} =  \frac{(-v)^{n}}{N^{n+2}} \sum_{b_{0},p,q}(-2)^{p}n_{b_{0}}(p,q) N^{b_{0}},
\end{align}

which has the following meaning: a given partition $(p,q,r)$ accounts for the diagrams with $p$ insertions of $L \otimes_{t} L^{\dagger}$, $q$ insertions of $L^{\dagger}L\otimes_{t} \mathbbm{1}$, and $n - p - q$ insertions of $\mathbbm{1}\otimes_{t} L^{\dagger}L$. A particular partition will have a total number of diagrams given by

\begin{align}
\left( { n \atop p} \right) \left( { n - p \atop q}\right).
\end{align}

However, since the insertions are not commutative, some of these can be distinct. This means there can be graphs with different topological index for the same partition. This is illustrated on a simple example in Fig.(\ref{fig:fourth-order-examples}). 

At a given order $n$, the number of connected components respects the bound $1 \le b_{0} \le n + 2$. The lower bound is assured since upon Wick contracting and taking the trace, one is guaranteed to have at least one connected component. The upper bound can be understood by the following argument: the identity ``propagator" consists of no insertions, which means that upon taking the trace we have two closed loops. An insertion on one of the loops corresponds, topologically, to cutting the loop. Thus at order $n$, with $n$ insertions bilinear in $L$ and $L^{\dagger}$, before Wick contraction there will be a total of $2n$ connected components. The rules for Wick contraction can be visualized by inserting a short tube into each gap, and requiring tubes to meet. This is just a pictorial way to say that the indices of one matrix must be contracted with the indices of only one other matrix; hence the tubes must connect to tubes. 

To understand the maximal possible $b_{0}$ for a given graph, we can proceed inductively and completely generally. The identity graph is two disconnected loops, and thus $b_{0} = 2$. A single insertion will produce two cuts, which will either be on the same loop or on different loops. Thus, after Wick contraction, we may have $b_{0} = 1$ or $b_{0} = 3$. A second insertion will now produce two more cuts, which may occur on the same loop or on different loops. Repeating the argument for the single insertion, we end up with the possibilities $b_{0} \in \{ 2, 4\}$. We may formally prove this by induction, but it is also clear by inspection that we can expect

\begin{align}
b_{0} &\in \{ 1,3, ..., n+2\}, \quad n \; {\rm odd}\\
b_{0} &\in \{ 2, 4, ..., n+2\}, \quad n \; {\rm even}
\end{align}

A new insertion can either join two loops, or split a single loop in two pieces. This recursive structure implies that to reach the maximal $b_{0}$ requires that one never join two loops. And since any number of insertions of the recycling term will result in a diagram that joins two loops, they do not participate in the leading order $N^{n+2}$ diagrams. Thus, we arrive at a proof of Lemma (\ref{lemma}) which is completely agnostic about the number of jump operators in the problem. 

 The topological expansion we have outlined in this appendix seems distinct from the classical topological expansion proposed for large $N$ gauge theories \cite{t1974planar} and matrix models \cite{brezin1978planar}. We speculate that the difference lies in the fact that our random fields are fourth-order tensors (as opposed to matrices), due to the tensor product structure of the Lindbladian. Since the proof of Lemma (\ref{lemma}) required nothing more sophisticated than Betti numbers, we did not seek a deeper connection to geometry by embedding our graphs in Riemann surfaces. However, this goal seems entirely attainable and could very well provide additional insights into the problem, possibly making the hermitian dissipator amenable to graphical enumeration techniques. 

We should also comment here that another natural setting in which the tensor product structure will arise is when there exist (spatial) locality constraints on an operator (see e.g. \cite{chan2018solution, kukuljan2016corner}). In this case, the spatial structure of the Hilbert space enforces the tensor product structure of operators (e.g. the Hamiltonian) acting in this space. It is possible that in this setting as well, one might benefit from the topological expansion discussed in this appendix. 

\begin{figure}[htbp!]
\centering
\includegraphics[scale=0.6]{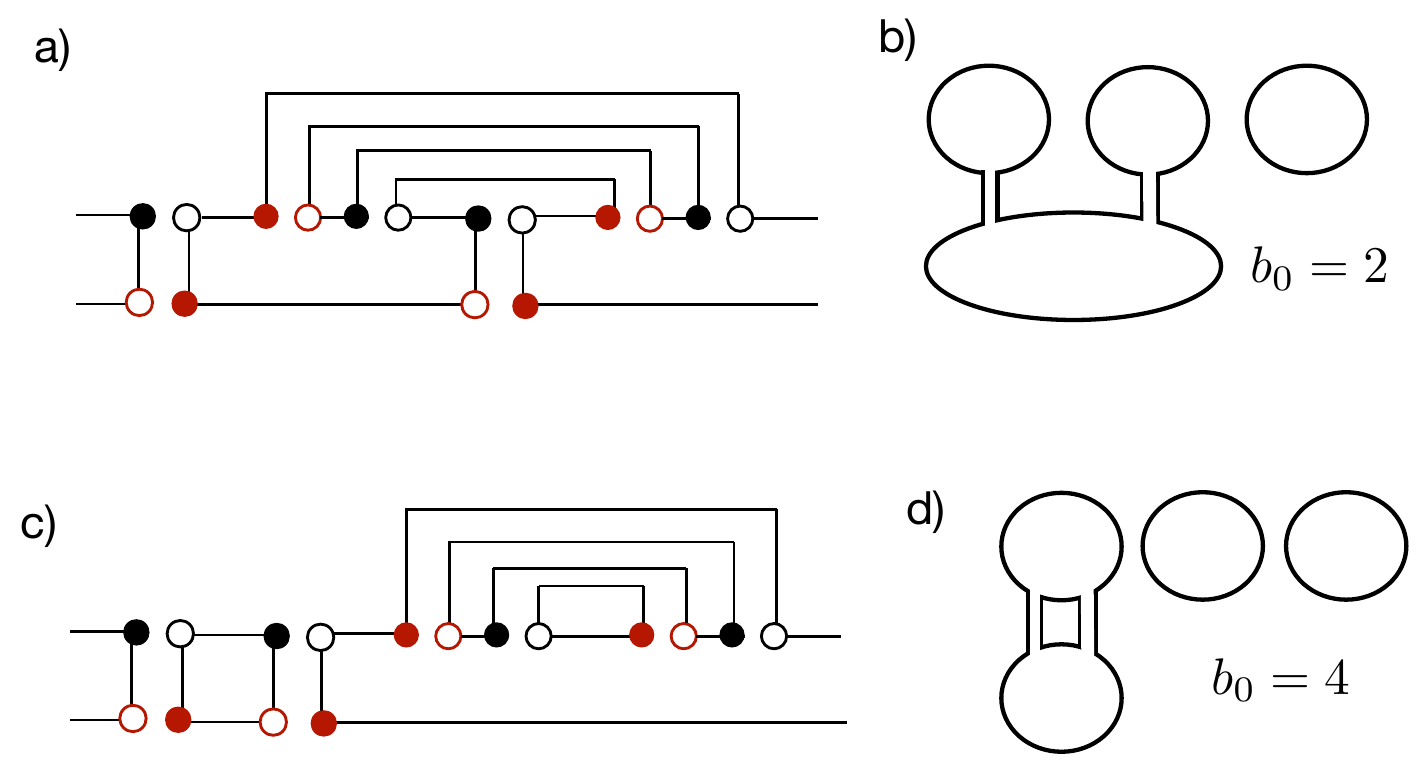}\\ 
\caption{Two diagrams appearing in the fourth moment of the dissipator in the same partition $(p,q) = (2,2)$: (a) shows a particular contraction, and (b) the topological graph obtained upon taking the trace. (c) is not equivalent to (a) under cyclic permutations of the trace, and has a larger $b_{0}$, as shown in (d).}
\la{fig:fourth-order-examples}
\end{figure}

\section{Random Normal Matrices: Correlation Functions and Exact Form Factor}\label{normal}

The joint probability distribution function of eigenvalues for a random normal matrix with Gaussian distributed entries follows the same law as the complex Ginibre ensemble, and is consequently of determinantal form. It is given by \cite{Zabrodin2004}
\begin{align}
P(z_{1}, ..., z_{N}) = \frac{1}{Z_{N}} \prod_{i<j} |z_{i} - z_{j}|^{2} \prod_{k = 1}^{N} e^{ - \frac{N}{v} |z_{k}|^{2}},	
\end{align}
where $Z_{N}$ is a normalization constant. All correlation functions follow from the kernel
\begin{align}
K(z_{1}, z_{2}) = \frac{N}{\pi v} \sum_{n = 0}^{N-1} \frac{1}{n!} \left( \frac{N z_{1	} \bar{z}_{2}}{v}\right)^{n} e^{ - \frac{N}{2v} (|z_{1}|^{2}  + |z_{2}|^{2})}.
\end{align}
For instance, the density is given by
\begin{align}
\langle \rho(z) \rangle = K(z, z) = \frac{N}{\pi v}\sum_{n = 0}^{N-1} \frac{1}{n!} \left( \frac{N |z|^{2}}{v}\right)^{n} e^{ - \frac{N}{v} |z|^{2}} 	, \quad \int d^{2} z \langle \rho(z) \rangle  = N. \label{density-normal}
\end{align}
This can be expressed more economically in terms of the incomplete Gamma function
\begin{align}
\langle \rho(z) \rangle = \frac{N}{\pi v} \frac{\Gamma(N, N |z|^{2}/v)}{\Gamma(N)},
\end{align}
which in the $N\to \infty$ limit describes a droplet with constant density inside the disk of radius $ R = \sqrt{v}$, and zero density outside. The two-point function is
\begin{align}
\langle \rho(z_{1}) \rho(z_{2}) \rangle = \langle \rho(z_{1}) \rangle \delta^{(2)}(z_{1} - z_{2}) + R(z_{1}, z_{2})	,
\end{align}
where
\begin{align}
R(z_{1}, z_{2}) = K(z_{1}, z_{1}) K(z_{2}, z_{2}) - | K(z_{1}, z_{2})|^{2}. \label{normal-2pt-fn}	
\end{align}
The simple dissipator with random normal jump operators has eigenvalues
\begin{align}
\zeta_{ij} = - |z_{i}|^{2} - |z_{j}|^{2} + 2 z_{i} \bar{z}_{j}.
\end{align}
In terms of the vectors on the 2D plane ${\bf r}_{1}$ and ${\bf r}_{2}$, this can be written
\begin{align}
\zeta_{ij} = x + i y = - | {\bf r}_{i} - {\bf r}_{j}|^{2} + 2 i |{\bf r}_{i} \times {\bf r}_{j}	|.
\end{align}
A simple geometric argument using the fact that $|r_{i}|\le \sqrt{v}$ implies that the support of the eigenvalues of the dissipator is described by a shifted circle law (see Fig.(\ref{fig:dissnormalevals}) )
\begin{align}
\frac{y^{2}}{4v^{2}} + \left( \frac{x}{2v} + 1\right)^{2} = 1,
\end{align}
which clearly shows that the spectral gap vanishes. 

The dissipative form factor is
\begin{align}
N^{2} F(t) &= \left\langle \sum_{i,j} e^{ - \left( |z_{i}|^{2} + |z_{j}|^{2} - 2 z_{i} \bar{z}_{j}\right)t}  \right\rangle,\\
& = \int d^{2} z_{1} \, d^{2} z_{2}\, \langle \rho(z_{1}) \rho(z_{2})\rangle 	e^{ - \left( |z_{1}|^{2} + |z_{2}|^{2} - 2 z_{1} \bar{z}_{2}\right)t}, \\
& = N + \int d^{2} z_{1} \, d^{2} z_{2}\, R(z_{1}, z_{2}) e^{ - \left( |z_{1}|^{2} + |z_{2}|^{2} - 2 z_{1} \bar{z}_{2}\right)t}.\label{ff-exact}
\end{align}
The exact result for finite $N$ (setting $v = 1$) is
\begin{align}
N^{2}F(t) = N + \frac{N^{2}}{t^{2}} \left( 1 - \left( 1 + \frac{t}{N}\right)^{-N}\right)^{2} - \frac{N^{2}}{t^{2}} \left( 1 - \left( \frac{1 + 2 t/N}{(1 + t/N)^{2}}\right)^{N} \right).	\label{exact-form-factor}
\end{align}
For $t <<N$, we may develop this expression as an asymptotic series for large $N$,
\begin{align}
N^{2}F(t) = &N^{2} \frac{(1 - e^{ - t})^{2}}{t^{2}} + N (e^{ - 2t} - e^{ -t}) \nonumber\\
&  + \frac{t}{12} \left[ (6 t - 8) e^{ - 2t} + (8 - 3t) e^{ - t} + 6 (t + 4) \right] + O(N^{-1})
\end{align}

Curiously, this indicates that for times much smaller than $t \sim N$, there exists quadratic {\it growth} at order $O(1)$, reminiscent of the sub-leading growth of the spectral form factor in closed chaotic systems \cite{Berry1985}, and similarly due to eigenvalue correlations. The evidence for the closing of the spectral gap appears early on, however, at order $N^{2}$, which displays long-time power-law decay. 

We split now the evaluation of the second integral in \eq{ff-exact} into two parts using \eq{normal-2pt-fn}. The first is
\begin{align}
&\int d^{2} z_{1} \, d^{2} z_{2}\, \langle \rho(z_{1}) \rangle \langle \rho(z_{2}) \rangle e^{ - \left( |z_{1}|^{2} + |z_{2}|^{2} - 2 z_{1} \bar{z}_{2}\right)t} = \left( \int d^{2} z \langle \rho(z) \rangle e^{ - |z|^{2} t} \right)^{2},\nonumber\\
& = N^{2}\left(\frac{1}{\pi} \sum_{n = 0}^{N-1}  \frac{1}{N(1+\tau)^{n+1}}\, \frac{1}{n!}\int d^{2} z'   (  |z'|^{2})^{n} e^{ -  |z'|^{2} } \right)^{2}, \quad \tau = \frac{t}{N},\nonumber\\
& = \frac{1}{ (1+\tau)^{2}}  \left(\sum_{n = 0}^{N-1}  \frac{1}{(1+\tau)^{n}} \right)^{2} = \frac{1}{\tau^{2}} \left( 1 - \frac{1}{(1+ \tau)^{N}}\right)^{2}.\label{integral1}
\end{align}
To get to the second expression in the first line, we use the fact that the density depends only on the absolute value of its argument. Upon expanding $\exp( 2 z_{1} \bar{z}_{2} t)$ in a power series, and performing the angular integration, only the first term in the expansion survives. The following steps follow by direct calculation using the explicit form of the density \eq{density-normal}.

The second part of the integral in \eq{ff-exact} is
\begin{align}
\int &d^{2} z_{1} d^{2} z_{2} | K(z_{1}, z_{2})|^{2} e^{ - t|z_{1}|^{2}  - t|z_{2}|^{2}	+ 2 t z_{1} \bar{z}_{2}}, \nonumber\\
&= \frac{1}{\pi^{2}}\sum_{p\le q} \frac{1}{p! q!}  \int d^{2}x\, d^{2}y\, ( x \bar{y})^{p} ( \bar{x} y)^{q} e^{ -  (1 + \tau) |x|^{2} - (1 + \tau) |y|^{2} +2  \tau  x \bar{y}}, \quad \tau = \frac{t}{N}, \quad (x,y) = \sqrt{N}(z_{1}, z_{2}),\nonumber\\
& = \frac{1}{\pi^{2}}\sum_{p\le q} \frac{1}{p! q!}  \int d^{2}x\, d^{2}y\, \frac{(2\tau)^{q-p}}{(q-p)!} |x|^{2q} |y|^{2q} e^{ -  (1 + \tau) |x|^{2} - (1 + \tau) |y|^{2} }
 =  \sum_{p\le q} \frac{1}{p!}   \frac{(2\tau)^{q-p}}{(q-p)!} \frac{(q!)}{(1 + \tau)^{2q + 2}},\nonumber\\
& = \frac{N^{2}}{t^{2}} \left( 1 - \frac{(1 + 2 \tau)^{N}}{ (1 + \tau)^{2N}}\right).\label{integral2}
\end{align}
This is also evaluated by using a power series expansion of $\exp(2 \tau x \bar{y})$ and noting that the angular integration will kill any terms with $p>q$, hence the restricted summation in the second line. The third line shows the terms in the integrand that do not vanish upon taking the angular integrals. The rest are straightforward calculations.

Combining \eq{integral1} and \eq{integral2} leads to the exact form factor \eq{exact-form-factor}.

\section{Random Liouville-von Neumann Generator}\label{sec:spec-LvN}

For easy reference, we discuss here the spectral properties of the Liouville-von Neumann operator when the Hamiltonian is a random matrix. These results are used in Sec.(\ref{sec:diss-gue}) to discuss the simple dissipator with GUE jump operators. The primary objects of interest are the resolvent and spectral form factor. We do not dwell on the details and refer to the literature for proofs and derivations (e.g. \cite{Mehta,forrester2010log}).

Let $S = H \otimes_{t} \mathbbm{1} - \mathbbm{1} \otimes_{t} H$, with $H$ a random matrix drawn from the Gaussian unitary ensemble. Studying the statistical properties of this operator gives us access to both the Liouville-von Neumann operator $\mathcal{C}_{H} =  - i S$ (\ref{von-Neumann}), and the simple dissipator with a hermitian jump operator $\mathcal{D}_{H} = - S^{2}$ (\ref{herm-dissipator}). We fix the variance of the matrix elements to be independent of the symmetry class of the ensemble, 
\begin{align}
\langle \bar{H}_{ij} H_{kl} \rangle = \frac{v}{N} \delta_{	ik}\delta_{jl}.
\end{align}
The eigenvalue density function is
\begin{align}
\rho(x) = \frac{1}{N} \sum_{i = 1}^{N} \delta(x - \lambda_{i}),
\end{align}
where $\lambda_{i}$ are the eigenvalues of $H$. The expectation value of the density in the large $N$ limit is a smooth function given by the semi-circle law
\begin{align}
\langle \rho(x) \rangle \to  \rho_{sc}(x) = \frac{1 }{2\pi v } \sqrt{4v - x^{2}}	, \quad  \int_{-2 \sqrt{v}}^{2\sqrt{v}} \rho_{sc}(x) dx = 1.
\end{align}
The variance $v$ determines the bandwidth to be $4 \sqrt{v}$. The $n^{th}$ trace moment of the Liouville-von Neumann operator is defined as the expectation value of $\tr \, S^{n}$. Since all odd powers will vanish due to the Gaussian statistics, we only have the even moments

\begin{align}
	\mu_{2k} \equiv  \frac{1}{N^{2}}\left\langle \tr \, S^{2k} \right \rangle  = \frac{1}{N^{2}} \left\langle \sum_{i,j} (\lambda_{i} - \lambda_{j})^{2k} \right\rangle. \label{moment-LvN}
\end{align}
This can be conveniently expressed in terms of the two-point eigenvalue density correlation function
\begin{align}
\mu_{2k} = \int \int dx dy \left\langle \rho(x) \rho(y) \right\rangle \, (x - y)^{2k}. 	\label{moment-correlator}
\end{align}
The trace moments enjoy a large $N$ expansion whose leading order $O(1)$ term is given by using the disconnected two-point correlation function in (\ref{moment-correlator}). The asymptotic value of the moment $M_{2k} = \lim_{N \to \infty} \mu_{2k}$ is then 
\begin{align}
M_{2k} = \int\int dx dy\, \rho_{sc}(x) \rho_{sc}(	y) (x - y)^{2k} dx dy = v^{k}\frac{4^{2k+1} \Gamma(k+1/2) \Gamma(k+3/2)}{\pi \Gamma(k+2)\Gamma(k+3)}. \label{moments}
\end{align}

We can confirm this by computing the first three exact moments at finite $N$
\begin{align}
\mu_{2} &= 2 v \left(1 - \frac{1}{N^{2}}\right)\\
\mu_{4} &= 10 v^{2} \left(1 - \frac{1}{N^{2}}\right)\\
\mu_{6} &= 2 v^{3} \left( 35 - \frac{30}{N^{2}}\right) \left( 1 - \frac{1}{N^{2}}\right)	
\end{align}
Note the common factor of $(N^{2} - 1)/N^{2}$ in these first few moments. This is due to the general structure of the ensemble averaged matrix moments discussed in Sec. (\ref{sec:gin-diss}).

The resolvent for the Liouville-von Neumann operator can be expressed as an asymptotic Laurent series for large $z$ with the coefficients given by the moments above

\begin{align}
G_{\mathcal{C}}(z) = \frac{1}{N^{2}}\left \langle \tr \, \frac{1}{z - \mathcal{C}_{H}}\right\rangle = \frac{1}{z}\sum_{k = 0}^{\infty}\frac{(-i)^{2k}\mu_{2k}}{z^{2k}},	
\end{align}
whereas the resolvent for the hermitian simple dissipator is
\begin{align}
G_{\mathcal{D}}(z) = \frac{1}{N^{2}}\left \langle \tr \, \frac{1}{z - \mathcal{D}_{H}^{2}}\right\rangle = \frac{1}{z}\sum_{k = 0}^{\infty}\frac{(-1)^{k}\mu_{2k}}{z^{k}}. 	
\end{align}
Substituting the exact large $N$ expression for the moments, we find that the resolvent can be written in terms of the Gauss hypergeometric function as

\begin{align}
G_{\mathcal{C}}(z) =\frac{z}{2 v} \left( 1 -\, {}_{2} F_1 \left( - \frac{1}{2}, \frac{1}{2}; 2; \frac{16 v}{z^{2}}\right) \right) ,\label{resolvent-lvn}
\end{align}
with the dissipator resolvent quoted in Eq.\eq{resolvent-dissipator}. We present a proof of this in Sec.(\ref{sec:diss-gue}). The large $N$ limit here has the unfortunate feature that it completely misses the $N$ exact zero eigenvalues of both $\mathcal{C}_{H}$ and $\mathcal{D}_{H}$, since these would appear as a subleading correction $\frac{1}{N z}$.

The Gauss hypergeometric function has singularities when its argument takes the values $0$, $1$ and $\infty$. Since the power series is convergent only when the argument $16 v/ z^{2}$ is outside the unit circle, analytical continuation inside the unit disk is required. To this end, we can employ the Pfaff transformation of the hypergeometric functions \cite{Yoshida1997,Koepf1998}
\begin{align}
{}_{2}F_{1}\left(a, b; c; - \frac{x}{1 - x}\right) = (1 - x)^{a} 	{}_{2}F_{1}(a, c-b; c; x),\label{pfaff}
\end{align}
which we use to evaluate the resolvent when $z$ is close to zero. To this end, let $a = -1/2$, $c = 2$, $b = 3/2$, and $x = 16 v/ z^{2}$. Then we may use the Pfaff transformation to get that as $x \to \infty$,
\begin{align}
  {}_{2} F_1 \left( - \frac{1}{2}, \frac{1}{2}; 2; x\right)  \approx   \frac{4}{3\pi} (1 - x)^{1/2},
\end{align}
which implies for the resolvent (\ref{resolvent-lvn})
\begin{align}
G(z)=  \frac{z}{2v} - \frac{1}{2v} \frac{4}{3\pi} \sqrt{ z^{2} - 16 v}   , \quad z \to 0.
\end{align}
From this we see that the eigenvalue density has finite support in the domain $[ - 4 \sqrt{v}, 4 \sqrt{v}]$. The total length of the branch cut gives the bandwidth of the eigenvalue distribution, which in this case is $8 \sqrt{v}$. As expected, this is twice the bandwidth of the spectrum of the Hamiltonian. 

The leading order contribution to the spectral form factor of the Liouville-von Neumann operator can also be obtained by the method of moments, although a direct integration using the semi-circle law is much simpler. Comparing these, we find two representations of the spectral form factor 
\begin{align}
\lim_{N \to \infty} \frac{1}{N^{2}} \left\langle e^{ t \mathcal{C}_{H}} \right\rangle = \left(\frac{J_{1}(2 \sqrt{v} t)}{\sqrt{v} t}\right)^{2} = {}_{1}F_{2}\left( \frac{3}{2}; 2, 3; - 4 v t^{2}\right).\label{sff-H}
\end{align}

The first expression, in terms of the Bessel function of the first kind, is very well known, while perhaps the latter, in terms of the generalized hypergeometric function, is less known. Using well-known asymptotics for the Bessel function one can verify the $t^{-3}$ power-law decay for the leading order contribution to the spectral form factor at long-times.

\section{Random Non-Hermitian Hamiltonian: Resolvent and Eigenvalue Density}\label{sec:nonhermitian-ham}

Here we use the Hermitization trick \cite{Feinberg1997,Janik1997PRE} (nicely reviewed in \cite{Feinberg2006,Janik1999a}) combined with a linearization procedure \cite{Belinschi2018} to obtain the resolvent for the non-Hermitian hamiltonian. Our results here reproduce those of \cite{Haake1992,Lehmann1995}, which were later significantly elaborated upon in \cite{Fyodorov1997a}. 

Let $H$ be a GUE matrix with variance $\langle |H_{ij}|^{2} \rangle = N^{-1}$, and $L$ be a complex Ginibre matrix with variance $\langle |L_{ij}|^{2}\rangle = \gamma / N$, where $N$ is the dimension of the matrices. Let $z$ be a complex coordinate (the argument of our Green's function), and $\eta$ an auxiliary complex variable introduced to analytically continue the Green's function ``off" the complex plane \cite{Feinberg1997}. Consider the Green's function obtained by inverting the following $4N\times 4N$ matrix
\begin{align}
{\bf G}^{-1} = \left( \begin{array}{cccc}
 \eta & \quad z + i H  & \quad 	0 &\quad L^{\dagger}\\
 \bar{z} - i H & \quad \eta & \quad L^{\dagger} & \quad 0\\
 L & \quad 0 & \quad \mathbbm{1} & \quad 0\\
 0 & \quad L & \quad 0 & \quad \mathbbm{1}
 \end{array}\right).
\end{align}
Let $A = z + i H - L^{\dagger}L$, and define 
\begin{align}
\mathcal{G} = \left( \begin{array}{cc} 
 	\eta & \quad A\\
 A^{\dagger} & \quad \eta
 	 \end{array}\right)^{-1} = \left( \begin{array}{cc} 
 	\frac{\eta}{\eta^{2} - A A^{\dagger}}  & \quad - A \frac{1}{\eta^{2} - A^{\dagger} A}   \\
- A^{\dagger} \frac{1}{\eta^{2} - A A^{\dagger}} & \quad \frac{\eta}{\eta^{2} - A^{\dagger} A}
 	 \end{array}\right) \equiv \left( \begin{array}{cc} 
 	G_{11}  & \quad G_{12}  \\
G_{21} & \quad G_{22}
 	 \end{array}\right).
\end{align}
Then the elements of the full Green's function here becomes
\begin{align}
{\bf G} = \left( \begin{array}{cccc}
G_{11} & \quad 	G_{12} & \quad - G_{12} L^{\dagger} & \quad - G_{11} L^{\dagger}\\
G_{21} & \quad G_{22} & \quad - G_{22}L^{\dagger} & \quad - G_{21}L^{\dagger}\\
- L G_{11}  & \quad - L G_{12} & \quad \mathbbm{1} + L G_{12}L^{\dagger} & \quad L G_{11} L^{\dagger}\\
- L G_{21} & \quad - L G_{22} & \quad L G_{22}L^{\dagger} &\quad \mathbbm{1} + L G_{21}L^{\dagger}
 \end{array}\right).
\end{align}

We are ultimately interested in obtaining an expression for $G_{21}$, which in the $\eta \to 0$ limit becomes the resolvent $G_{21} = A^{-1}$ of the non-Hermitian Hamiltonian.

By symmetry arguments, we can see that the off-diagonal blocks of ${\bf G}$ will be zero after ensemble averaging. Since the matrix is linear in Gaussian fields, the self-consistent Born approximation, which consists of keeping only the second cumulant in the self-energy, must be exact in the large $N$ limit. In order to do this, we write
\begin{align}
{\bf G}^{-1} = {\bf G}_{0}^{-1} - {\bf H},
\end{align}
where
\begin{align}
{\bf G}_{0}^{-1} = \left( \begin{array}{cccc}
 \eta & \quad z  & \quad 	0 &\quad 0\\
 \bar{z}  & \quad \eta & \quad 0 & \quad 0\\
 0 & \quad 0 & \quad \mathbbm{1} & \quad 0\\
 0 & \quad 0 & \quad 0 & \quad \mathbbm{1}
 \end{array}\right), \quad {\bf H} = \left( \begin{array}{cccc}
 0 & \quad  -i H  & \quad 	0 &\quad -L^{\dagger}\\
  i H & \quad 0 & \quad -L^{\dagger} & \quad 0\\
 -L & \quad 0 & \quad 0 & \quad 0\\
 0 & \quad -L & \quad 0 & \quad 0
 \end{array}\right).
\end{align}
The self-energy functional is then
\begin{align}
\Sigma[{\bf F}] = \left\langle 	{\bf H} {\bf F} {\bf H}\right\rangle .
\end{align}
This will involve expectation values of the following form
\begin{align}
\langle H f H \rangle &= 	\frac{1}{N} \tr(f) \mathbbm{1}, \quad \langle L f L^{\dagger} \rangle  = \frac{\gamma}{N} \tr (f) \mathbbm{1}.
\end{align}
Defining
\begin{align}
g_{ab} = \frac{1}{N} \tr\, {\bf G}_{ab}	,
\end{align}
we wind up with the self-energy
\begin{align}
\Sigma[{\bf G}] = \left( \begin{array}{cccc}
 g_{22} + \gamma g_{43} & \quad - g_{21} + \gamma g_{44} & \quad 0 & \quad 0\\
 - g_{12} + \gamma g_{33} & \quad g_{11} + \gamma g_{34} & \quad 0 & \quad 0\\
 0 & \quad 0 & \quad \gamma g_{12} & \quad \gamma g_{11}\\
 0 & \quad 0 & \quad \gamma g_{22} & \quad \gamma g_{21}	
 \end{array}\right).
\end{align}
Note that each block is actually a matrix proportional to the $N\times N$ identity matrix; for ease of notation, we write only the proportionality constants, which are functions of $(z, \bar{z})$.

By inspection, we see that upon disorder averaging, we must have
\begin{align}
g_{11} = g_{22}, \quad g_{12} = \bar{g}_{21},\quad  g_{33}  = \bar{g}_{44}	, \quad g_{34} = g_{43},
\end{align}
which simplifies the self-energy considerably
\begin{align}
\Sigma[{\bf G}] = \left( \begin{array}{cccc}
 g_{11} + \gamma g_{34} & \quad - g_{21} + \gamma g_{44} & \quad 0 & \quad 0\\
 - \bar{g}_{21} + \gamma \bar{g}_{44} & \quad g_{11} + \gamma g_{34} & \quad 0 & \quad 0\\
 0 & \quad 0 & \quad \gamma \bar{g}_{21} & \quad \gamma g_{11}\\
 0 & \quad 0 & \quad \gamma g_{11} & \quad \gamma g_{21}	
 \end{array}\right).
\end{align}
When solving the Dyson equation for the Green's function with this self-energy functional in the self-consistent Born approximation $\langle {\bf G}\rangle^{-1} = {\bf G}_{0}^{-1} - \Sigma[\langle {\bf G}\rangle ]$, we can take $\eta = 0$ and invert the following matrix
\begin{align}
\langle {\bf G}\rangle^{-1} = 	\left( \begin{array}{cccc}
 -g_{11} - \gamma g_{34} & \quad  z+g_{21} - \gamma g_{44} & \quad 0 & \quad 0\\
\bar{z}+ \bar{g}_{21} - \gamma \bar{g}_{44} & \quad -g_{11} - \gamma g_{34} & \quad 0 & \quad 0\\
 0 & \quad 0 & \quad 1-\gamma \bar{g}_{21} & \quad -\gamma g_{11}\\
 0 & \quad 0 & \quad -\gamma g_{11} & \quad 1-\gamma g_{21}	
 \end{array}\right).
\end{align}
This implies
\begin{align}
\left( \begin{array}{cc}
 \bar{g}_{44} & g_{34}\\
 g_{34} & g_{44}
 \end{array}\right)
 = \frac{1}{|1 - \gamma g_{21}|^{2} - \gamma^{2} g_{11}^{2}} \left( \begin{array}{cc}
 1 - \gamma g_{21} & \gamma g_{11}\\
 \gamma g_{11} & 1 - \gamma \bar{g}_{21}	
 \end{array}\right),
\end{align}
which gives us
\begin{align}
g_{44} = \frac{1 - \gamma \bar{g}_{21}}{|1 - \gamma g_{21}|^{2} - \gamma^{2} g_{11}^{2}}, \quad g_{34} = \frac{\gamma g_{11}}{	|1 - \gamma g_{21}|^{2} - \gamma^{2} g_{11}^{2}}.
\end{align}
Similarly, for the upper block we get the equations 
\begin{align}
g_{11} &= \frac{g_{11} + \gamma g_{34}}{	 | z + g_{21} - \gamma g_{44}|^{2} - (g_{11} + \gamma g_{34})^{2} },\\
g_{21}& = \frac{\bar{z} + \bar{g}_{21} - \gamma \bar{g}_{44}}{	 | z + g_{21} - \gamma g_{44}|^{2} - (g_{11} + \gamma g_{34})^{2} }.
\end{align}
To find the solution inside the support of the eigenvalue density, we first consider the solution with $g_{11}\ne 0$. At this point, we have a straightforward, if extremely tedious, algebraic problem on our hands. Its solution can be found by hand or using symbolic manipulation software (e.g. Mathematica), resulting in the sought after trace of the resolvent
\begin{align}
g_{21} = \frac{1}{2} \left( - \frac{1}{x} + \frac{\gamma}{1 + \gamma x} + \frac{1}{\gamma}\right) - i \frac{y}{2}	,
\end{align}
from which follows the mean density 
\begin{align}
\rho(x,y) = \frac{1}{\pi} \partial_{\bar{z}} g_{21} = \frac{1}{4\pi} \left[ \frac{1}{x^{2}} - \frac{\gamma^{2}}{(1 + \gamma x)^{2}} + 1 \right]	,
\end{align}
which is independent of the horizontal direction $y$. The resolvent for the singular values $g_{11}$ can be written 
\begin{align}
\gamma^{2} g_{11}^{2} = | 1 - \gamma g_{21}|^{2} - \frac{\gamma}{\bar{z} + \bar{g}_{21} - g_{21}} =  \frac{1}{4} \left[ \frac{\gamma}{x} - \frac{\gamma^{2}}{ (1 + \gamma x)}  + 1 \right]^{2} + \frac{\gamma^{2} y^{2}}{4} - \frac{\gamma}{x}.
\end{align}
The boundary of the eigenvalue support is described parametrically by the condition $g_{11} = 0$, giving 
\begin{align}
 \left[ \frac{1}{x} - \frac{\gamma}{ (1 + \gamma x)}  + \frac{1}{\gamma} \right]^{2}  - \frac{4}{ \gamma x}+y^{2} = 0,
\end{align}

or 
\begin{align}
y(x)^{2} = \frac{4}{\gamma x} - 	\left[ \frac{1}{x} - \frac{\gamma}{ (1 + \gamma x)}  + \frac{1}{\gamma} \right]^{2}.\label{y-support}
\end{align}

\section{Non-Hermitian Hamiltonian with Multiple Jump Operators}\label{sec:lind-m}

Here we provide an argument for the generalization of the spectral support curve (\ref{y-support}) for a non-Hermitian Hamiltonian with multiple jump operators
\begin{align}
 -i H + \gamma  \sum_{a = 1}^{m} L_{a}^{\dagger}L_{a}, \label{nh-m}
\end{align}
where $L_{a}$ are independent random complex Ginibre matrices whose elements have variance $1/N$. By following the steps taken in the replica approach used in \cite{Haake1992}, it is clear that the curve describing the support of the eigenvalues is simply modified to read
\begin{align}
y(x)^{2} = \frac{4m}{\gamma x} - 	\left[ \frac{m}{x} - \frac{\gamma}{ (1 + \gamma x)}  + \frac{1}{\gamma} \right]^{2}.\label{y-support-m}
\end{align}
In \cite{Haake1992}, $mN$ is the number of scattering channels, and $m<1$ is implicitly assumed in their work. However, this assumption does not appear necessary, and it is simple to see that it should generalize to our case in which $m \in \mathbbm{N}$ is an integer greater than one.

At small $\gamma$, we can drop the second term in brackets in (\ref{y-support-m}), and solve 
\begin{align}
\frac{4 m}{\gamma x} \approx  \left( \frac{m}{x} + \frac{1}{\gamma}\right)^{2}. 
\end{align}
The unique solution is $x_{min} = m \gamma$. However, the notable difference between $m = 1$ is seen at large $\gamma$. This is perhaps not surprising given the fact that in the $\gamma \to \infty$ limit, the spectrum remains gapped, according to (\ref{multi-diss-density-m}). To proceed, we can utilize the scaling ansatz again $x_{min} \sim x_{0} \gamma^{\alpha}$, and assuming $\alpha >0$, we are left to solve

\begin{align}
	\frac{4 m}{x_{0} \gamma^{\alpha+1}} = \left( \frac{(m-1)}{x_{0} \gamma^{\alpha}} + \frac{1}{\gamma} \right)^{2}.
\end{align}
Self-consistency requires $\alpha + 1  = {\rm min} ( 2 , 2 \alpha)$, which is only compatible with $\alpha = 1$. Now plugging this back in we can solve for the prefactor
\begin{align}
\frac{4 m}{x_{0}} = \left( \frac{m - 1}{x_{0}} + 1 \right)^{2},
\end{align}
which has the solutions $x_{0} = (1 \pm \sqrt{m})^{2}$. The correct solution comes by requiring $x_{0}$ vanish at $m = 1$. Summarizing, we have for the asympototic behavior of the spectral gap of (\ref{nh-m})

\begin{align}
x_{min} = \begin{cases}
 \gamma m , \quad &\gamma \to 0,\\
\gamma (1 - \sqrt{m})^{2}, \quad &\gamma \to \infty.\\
\end{cases}	
\end{align}
Using that the spectral gap of the Lindbladian is $\langle \Delta \rangle = 2 x_{min}$, we recover the results found recently in \cite{sa2019spectral}. Note that the spectral gap for $m>1$ appears to coincide for large $\gamma$ with the pure dissipator.  

\bibliographystyle{unsrt}
\bibliography{random-gkls.bib}

\end{document}